\documentclass[nofootinbib,notitlepage,superscriptaddress,10pt,aps,pra,twocolumn]{revtex4-2}

\usepackage{xcolor}
\usepackage[utf8x]{inputenc}
\usepackage{amsmath}
\usepackage{amssymb}
\usepackage{graphicx}
\usepackage{float}
\usepackage[normalem]{ulem}
\usepackage{epsf,epsfig}
\usepackage{psfrag}
\usepackage{color}
\usepackage{soul}

\makeatletter

\begin{document}

\title{Asymmetry in momentum space: restoring CPT invariance of $\kappa$-field theory}

\author{Tadeusz Adach}
\affiliation{University of Wroc\l{}aw, Faculty of Physics and Astronomy, pl.\ M.\ Borna 9, 50-204
Wroc\l{}aw, Poland}
\author{Andrea Bevilacqua}
\affiliation{National Centre for Nuclear Research, ul. Pasteura 7, 02-093 Warsaw, Poland}
\author{Jerzy Kowalski-Glikman}
\affiliation{National Centre for Nuclear Research, ul. Pasteura 7, 02-093 Warsaw, Poland}
\affiliation{University of Wroc\l{}aw, Faculty of Physics and Astronomy, pl.\ M.\ Borna 9, 50-204
Wroc\l{}aw, Poland}
\author{Giacomo Rosati}
\affiliation{Dipartimento di Matematica, Università di Cagliari, via Ospedale 72, 09124 Cagliari, Italy}
\affiliation{Istituto Nazionale di Fisica Nucleare, Sezione di Cagliari,
Cittadella Universitaria, 09042 Monserrato, Italy}

\begin{abstract}

The positive and negative energy modes of a field theory in $\kappa$-Minkowski/$\kappa$-Poincar\'e noncommutative spacetime have very different symmetry properties.
This can be understood geometrically by considering that they span two distinct sectors of a curved momentum space.
By performing an explicit direct computation of the relativistic Noether charges and their algebra within the canonical formalism, we identify a striking consequence of this asymmetry in momentum space: charge conjugation and Poincar\'e invariance are incompatible.
We then notice how the structure of momentum space suggests that time reversal could be deformed so that the overall CPT-invariance is restored.
We prove that this new proposal works by studying the transformation properties under deformed discrete symmetries of the new relativistic charges.

\end{abstract}

\maketitle

\section{Introduction}\label{sec:intro}

The $\kappa$-field theory~\cite{LukKosMaskfield,GACmichelekfield,JurekDaszkImilkfield2004,GACnoether2007,FreJurekNowkfield2,LukDaszkWorkfield,MicheleJurekLorentz,kDiscrete1} (the theory of fields in $\kappa$-Minkowski noncommutative spacetime~\cite{MajidRuegg} with $\kappa$-Poincar\'e deformed relativistic symmetries~\cite{LukRueggNowTolkPoincare,LukNowRueggkPpoinc}) has played a prominent role in the context of quantum gravity phenomenology, as it provides a fruitful framework in which to test the properties of Planck-scale deformed relativistic symmetries, characterizing effective models of quantum spacetimes~\cite{GACreview,COSTreview}.

It is now well understood that in a theory of fields living in a noncommutative spacetime of Lie algebra type, as is the case for $\kappa$-Minkowski, the momentum space is a group manifold, which in general has curvature~\cite{JurekdeSitter1,JurekdeSitter2,micheleAnatomy,kDiscrete1}.
One consequence of the curvature of momentum space~\cite{micheleAnatomy} is that in such a theory positive and negative frequency modes are distinct, in the sense that, apart from being characterized by a different sign in the ``time-frequency" part (the energy of the field mode), the geometry of momentum space for (on-shell) positive and negative frequency modes is fundamentally different (see Sec.~\ref{sec:momentum space}).

The aim of this work is to investigate the consequences of this asymmetry for the consistency of continuous and discrete relativistic symmetries.

The properties of $\kappa$-momentum space have been studied in several previous works, which have shed light on some aspects of its symmetries.
One important step forward has been achieved in \cite{MicheleJurekLorentz} in understanding how the positive/negative frequency sectors of momentum space transform under (deformed) Lorentz transformations. The transformation properties defined in \cite{MicheleJurekLorentz} rely on the so-called ``antipode" map (in the sense of Hopf-algebra) of the $\kappa$-Poincar\'e algebra.
The crucial observation of~\cite{MicheleJurekLorentz} can be rephrased in these terms as follows (see also~\cite{JurekMicheleDiscrete}): if the momenta of a mode in the positive frequency sector transform with the action of the Lorentz generators (of $\kappa$-Poincar\'e) algebra, for the corresponding mode in the negative frequency sector -- whose momenta are the antipode of the ones of the positive frequency mode -- they transform with the antipode of the Lorentz generator.

The results of \cite{MicheleJurekLorentz} were at the basis of a series of works~\cite{JurekMicheleDiscrete,kDiscrete1,JurekAndreaDiscrete} aiming to study the discrete symmetries of a scalar field theory with $\kappa$-momentum space, in which some of the authors of this work were also involved.

One of the aims of these previous works was to formulate an action for a scalar field explicitly invariant under charge conjugation ${\cal C}$, intended (as usual) as an exchange between positive and negative frequency modes, which, after quantization, corresponds to an exchange between particles and antiparticles.
One puzzling aspect emerging in~\cite{kDiscrete1,JurekAndreaDiscrete} is that even if one starts with an apparently ${\cal C}$-invariant Lagrangian, the ${\cal C}$-invariance seems to be broken at the level of Noether charges. Specifically, it was observed that the Lorentz charges obtained from the action in~\cite{kDiscrete1,JurekAndreaDiscrete} are not ${\cal C}$-invariant.

To investigate this problem, the first objective of the present work is to derive the whole set of relativistic conserved charges (in particular, the Lorentz charges) from a direct calculation based on the canonical Noether analysis of the $\kappa$-field action. 
In most of the past works,  only the translational charges were derived directly with the canonical formalism, while the Lorentz charges were obtained through a ``covariant phase space" approach~\cite{kDiscrete1,JurekAndreaDiscrete} that dramatically reduced technical difficulties (an  exception being \cite{FreJurekNowkfield2}, which however considered a model different from the one we discuss here).
That approach, even though in principle rather efficient in treating Noether analysis, obscured some of the underlying assumptions needed to make it work consistently. 
It is only through a direct calculation based on the canonical formalism that one can shed light on these hidden features.

The noncommutative structure of the theory allows indeed to consider different ordering prescriptions for the Lagrangian, all reducing to the same classical limit.
The $\cal C$-invariance of the model, however, depends on the choice of the ordering prescription.
Aside from the calculation of the translational charges, we thus perform the explicit direct calculation of the Noether charges associated with Lorentz transformations for the $\kappa$-field action, studying how the structure of the charges and their symmetry properties depend on the ordering of the terms in the action. 
We then obtain the symplectic structure associated with the action and compute the full Poincaré algebra.

This is the first result of the present work.
In particular, the comparison between the fully direct approach and the covariant phase space approach brings to light the role that boundary terms play for the (incompatible) symmetry properties of the deformed action under continuous and discrete transformations, a feature that was previously missing in the literature.
Specifically, one finds that even if one starts with a Lagrangian whose combination of terms is ordered so that the starting model is $\cal C$-invariant, the covariant phase space approach of~\cite{kDiscrete1,JurekAndreaDiscrete} automatically modifies the starting Lagrangian by (implicitly) adding boundary terms in order to make it consistent with the Poincar\'e algebra, at the cost of breaking the $\cal{C}$ invariance of the starting model. In other words, the covariant phase space approach picks up only a subclass of the possible Lagrangians that could be admissible in principle by the different ordering prescriptions, that indeed turn out to differ by boundary terms\footnote{This implies also that Lagrangians corresponding to different ordering prescriptions can be traced back to the same action.
The discrepancy between the $\kappa$-field action and Lagrangian was discussed preliminarily in~\cite{AdachPoS}.}.
On the other hand, the direct canonical approach keeps track of all the boundary terms, and at the end allows one to obtain conserved charges for all classes of models, including those that are $\cal C$-invariant but break the Poincar\'e algebra.

On the conceptual level, focusing on the geometry of momentum space, the main consideration underlying our analysis is that if charge conjugation is defined as a transformation from the positive to the negative frequency sectors (or vice versa) of momentum space, then it is hard to expect that such a symmetry will survive in a theory where these sectors are distinct.
The results of our investigation confirm this expectation.
More specifically, one finds that there is no way to preserve simultaneously charge conjugation and Poincaré symmetries.

Our thesis is that the asymmetry in $\kappa$-momentum space is responsible for the fact that a field theory invariant under $\kappa$-deformed relativistic symmetries cannot be ${\cal C}$-invariant. Even if this aspect might not be particularly concerning, unlike charge conjugation, ${\cal C}{\cal P}{\cal T}$ symmetry is widely regarded as a fundamental feature of all known relativistic (Lorentz-invariant) field theories. While ${\cal C}{\cal P}{\cal T}$ violation is naturally expected in frameworks where Lorentz invariance is explicitly broken, it would be conceptually troubling if ${\cal C}{\cal P}{\cal T}$ symmetry were violated in a theory that still admits a full set of relativistic symmetries, even if deformed.

The second result of the present work is that there is a natural -- albeit nontrivial -- way in which ${\cal C}{\cal P}{\cal T}$-invariance can be restored in our noncommutative framework.
Besides charge conjugation and parity, we propose a new definition of the action of time reversal on fields, motivated both by the algebraic properties of $\kappa$-momentum space and by its geometrical structure. 
We study the action of the whole set of discrete symmetries on fields, and show how, with our new definition of time reversal, ${\cal C}{\cal P}{\cal T}$ symmetry leaves both the $\kappa$-field action and the Poincar\'e invariant relativistic charges invariant.
We thus finally prove that, with our new proposal, while the Poincar\'e invariant $\kappa$-field models cannot be $\cal C$ invariant, they can be naturally invariant under ${\cal C}{\cal P}{\cal T}$.

Restoring ${\cal C}{\cal P}{\cal T}$-invariance is possible due to the flexibility inherent in defining discrete transformations within the deformed framework..
In particular, we assume a definition of time reversal ${\cal T}$ different from the one used in the previous works~\cite{kDiscrete1,JurekAndreaDiscrete}.
However, we notice that both definitions are technically plausible, since they both reduce to the standard action of ${\cal T}$ in the undeformed case, even if they have significantly different conceptual, and also phenomenological, implications.
We are going to comment on this point shortly in the conclusions.

The paper is organized as follows:

In section~\ref{sec:momentum space} we briefly recall the geometry of $\kappa$-momentum space.
In section~\ref{sec:complex field} we define the scalar field and introduce its Lagrangian and action.
In section~\ref{sec:complex noether}, we perform the Noether analysis for the Lagrangian introduced in sec.~\ref{sec:complex field} and calculate the conserved charges associated with Poincar\'e transformations and their algebra. 
In section~\ref{sec:chargeConjugation}, we introduce a definition of charge conjugation and discuss its relationship with the theory described in the previous sections. We also show how the incompatibility between ${\cal C}$ and Poincar\'e invariance emerges, how it is related to the addition of boundary terms to the action, and we clarify the aspects and implications of previous attempts to compensate for it by comparing the direct Noether computations presented here with previous covariant phase space computations.
In section~\ref{sec:parity and time reversal} we propose a new definition for the remaining discrete transformations and their combination (${\cal C}{\cal P}{\cal T}$).
In section~\ref{sec:CPTinvariance}, we demonstrate that the action of sec.~\ref{sec:complex field} and its conserved charges (sec.~\ref{sec:complex noether}) are in fact invariant under this transformation.
Finally, in section~\ref{sec:conclusions} we discuss the results of our analysis and comment shortly on possible implications of the new definition of discrete symmetries.

\section{$\kappa$-momentum space}
\label{sec:momentum space}

We present here a concise description of the main properties of the $\kappa$-momentum space. Other basic features of the structure of its symmetries are reported in App.~\ref{app:kPoincare}.

Considering $\kappa$-Minkowski noncommutative spacetime, defined by the relations
\begin{equation}
[\hat{x}^0,\hat{x}^j] = \frac{i}{\kappa}\hat{x}^j\ , \qquad [\hat{x}^j,\hat{x}^k] = 0\ ,
\end{equation}

\onecolumngrid

\begin{figure}[h]
\begin{centering}
\includegraphics[scale=0.45]{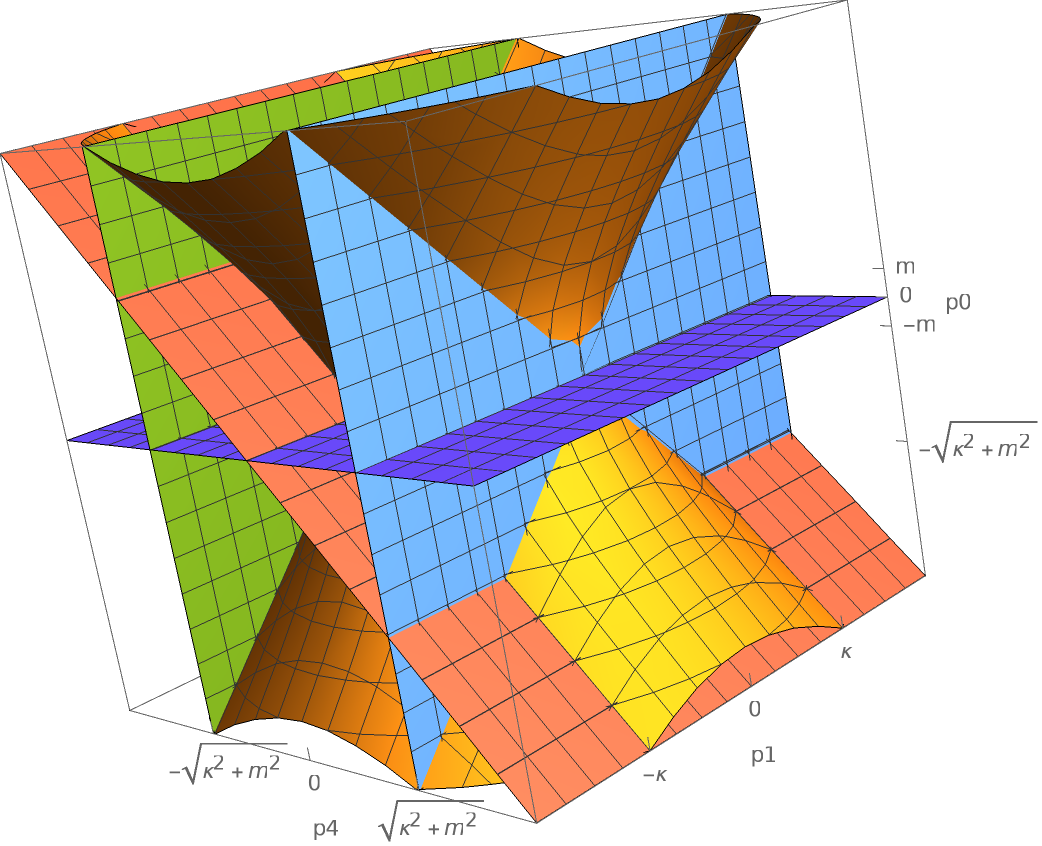}
~~~~~~ \includegraphics[scale=0.37]
{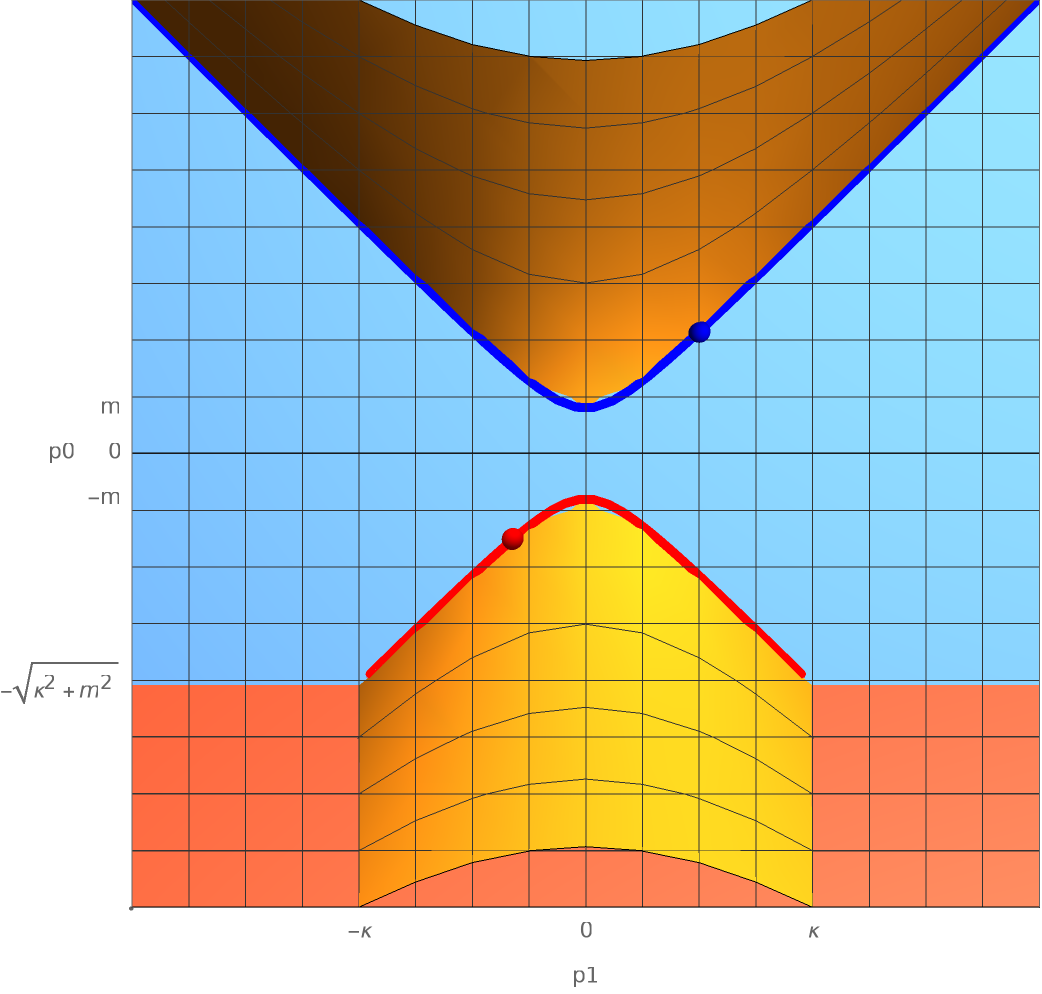}
\caption{\label{fig:-momentum-space}(1+1)D depiction of $\kappa$-momentum
space. The portion of the hyperboloid covered by the momentum space coordinates is the (yellow) one over the $p_+=p_0+p_4$ (orange) plane and for positive $p_4$ (l.h.s.). The on-shell orbits are the one sectioned by the plane $p_4=\sqrt{m^{2}+\kappa^{2}}$, respectively the blue (positive frequency) and red (negative frequency) curves on the r.h.s.. The blue and red dots depict two particular on-shell modes connected by the antipode map.}
\par\end{centering}
\end{figure}
\twocolumngrid
\noindent
the time-ordered noncommutative plane waves
\begin{equation}
\label{noncomm plane wave}
\hat{e}_k(\hat{x}) = e^{ik_j \hat{x}^j} e^{ik_0 \hat{x}^0}
\end{equation}
span a group manifold that constitutes the $AN_3$ space~\cite{JurekdeSitter1,JurekdeSitter2,FreJurekNowkfield2,kDiscrete1}, covering half of de Sitter space.
In terms of 5D embedding (momentum-space) coordinates 
\begin{equation}
\begin{split} & p_{0}=\kappa\sinh\left(k_{0}/\kappa\right)+\frac{1}{2\kappa}e^{k_{0}/\kappa}{\bf k}^{2}\\
 & p_{j}=e^{k_{0}/\kappa}k_{j}\\
 & p_{4}=\kappa\cosh\left(k_{0}/\kappa\right)-\frac{1}{2\kappa}e^{k_{0}/\kappa}{\bf k}^{2}
\end{split}
\label{embbedding map}
\end{equation}
this space can be identified with the de Sitter hyperboloid
\begin{equation}
p_{0}^{2}-{\bf p}^{2}-p_{4}^{2}=-\kappa^{2}
\end{equation}
restricted to the portion
\begin{equation}
\Delta_+(p)=p_{0}+p_{4}\equiv p_+>0\ ,\qquad p_{4}>0\ .\label{portionDeSitter}
\end{equation}
An on-shell ``positive energy mode'' (blue curve in Fig. \ref{fig:-momentum-space})
is described by the constraint
\begin{equation}
p_{0}=\omega_{{\bf p}}=\sqrt{m^{2}+{\bf p}^{2}}\ ,
\end{equation}
while a ``negative energy mode'' (red curve in Fig. \ref{fig:-momentum-space})
is obtained through the antipode map, so that it has energy and momentum
\begin{equation}
\begin{gathered}
\omega_{{\bf p}}\rightarrow 
S\left(p_{0}\right)\Big|_{p_{0}=\omega_{{\bf p}}}
=-\omega_{{\bf p}}+\frac{{\bf p}^{2}}{p_+}\\
{\bf p}\rightarrow
S\left({\bf p}\right)=-\frac{\kappa}{p_+}{\bf p}\ ,
\end{gathered}
\end{equation}
where $p_+=\Delta_{+}\left({\bf p}\right)$ is also on shell:
\begin{equation}
p_+=\omega_{{\bf p}}+p_{4}\left({\bf p}\right)\:,\qquad p_{4}\left({\bf p}\right)=\sqrt{m^{2}+\kappa^{2}}\ .
\end{equation}

The asymmetry of the momentum space is manifest: the on-shell orbits for the positive energy are unbounded, while the orbits of the negative energy modes are bounded by the values
\begin{equation}
S\left(p_{0}\right)\! \Big|_{p_{0}=\omega_{{\bf p}}} \!\!\! \in\left(-\sqrt{m^{2}+\kappa^{2}},-m\right]\ \!,\quad S \left({\bf p}\right)\in\left(-\kappa,\kappa\right)\ .
\end{equation}

\section{Complex field in $\kappa$-Minkowski}
\label{sec:complex field}

To study the properties of fields in noncommutative spacetime, and in particular their (noncommutative) Fourier transform, the main tool is provided by the Weyl map~\cite{AgoLizziWeyl2002,AgoGACdanreaHopf2004}. The Weyl map makes it possible to describe the noncommutative fields in terms of standard functions, where the noncommutative product is represented in terms of a $\star$-product, encoding an infinite series of derivatives on the fields.
The corresponding noncommutative Fourier transform then establishes a link between the field and its (curved) momentum space.

Given a noncommutative spacetime, the choice of Weyl map is not unique. We choose here to work with the Weyl map that was introduced in~\cite{FreJurekNowkfield2} for $\kappa$-Minkowski (see e.g.~\cite{AgoGACdanreaHopf2004,GACnoether2007} for different choices of Weyl maps), and which we denote here as the ``embedding'' Weyl map
\begin{equation}
\hat{e}_{k}\left(\hat{x}\right)={\cal W}\left(e^{ip_{\mu}\left(k\right)x^{\mu}}\right)\ .\label{embeddingWeylMap}
\end{equation}
This particular choice has the property that noncommutative derivatives emerging from the five-dimensional differential calculus\footnote{It is also possible to construct a 4D differential calculus in $\kappa$-Minkowski~\cite{OecklDiffCalc,RosatikDiff}. This is at the basis of the works~\cite{AgoGACdanreaHopf2004,GACnoether2007}.} introduced in~\cite{SitarzDiffCalc} (see~\cite{FlaviokDiff} for the properties of the calculus and their application to field theory) are mapped to standard derivatives on functions of coordinates, which in turn correspond to the embedding momenta defined in Sec.~\ref{sec:momentum space} (details of the properties of the calculus are reported in App.~\ref{app:calculus}).
This is the choice of Weyl map that was also used for the works~\cite{kDiscrete1,JurekAndreaDiscrete}, where the discrete symmetry sector of the theory was studied.

Given~(\ref{embeddingWeylMap}), an on-shell scalar field in $\kappa$-Minkowski can be defined~\cite{kDiscrete1} through the (inverse) noncommutative Fourier transform
\begin{equation}
\phi\left(x\right)=\int d\mu \left(p\left(k\right)\right)\delta\left(p_{\mu}p^{\mu}-m^{2}\right)\tilde{\phi}\left(p\right){\cal W}^{-1}\left(\hat{e}_k\left(\hat{x}\right)\right),\label{eq-field-fourier}
\end{equation}
where
\begin{equation}
d\mu \left(p \right) = 2\kappa d^{5}p\ \delta\left(p_{0}^{2}-{\bf p}^{2}-p_{4}^{2}+\kappa^{2}\right)\theta\left(p_{+}\right)\theta\left(p_{4}\right)
\end{equation}
is the left-invariant Haar measure on $AN_3$.
The on-shell mode decomposition \eqref{eq-field-fourier} can be reformulated~\cite{kDiscrete1} as
\begin{equation}\begin{split}
\label{on-shell field}
\phi(x)=&\int\frac{d^3p}{2\omega_\mathbf{p}p_4/\kappa}\tilde{\phi}(\omega_\mathbf{p},\mathbf{p})e^{i(\omega_\mathbf{p}t-\mathbf{p}\cdot\mathbf{x})}\\
&+\int\frac{d^3S(p)}{2\omega_{S(\mathbf{p})}p_4/\kappa}\tilde{\phi}(-\omega_{S(\mathbf{p})},S(\mathbf{p}))e^{i(-\omega_{S(\mathbf{p})}t-S(\mathbf{p})\cdot\mathbf{x})}\\
=&\int\frac{d^3p}{2\omega_\mathbf{p}p_4/\kappa}\left[\tilde{\phi}(\omega_\mathbf{p},\mathbf{p})e^{i(\omega_\mathbf{p}t-\mathbf{p}\cdot\mathbf{x})}\right.\\
&+\left.\frac{\kappa^3}{p_+^3}\tilde{\phi}(-\omega_{S(\mathbf{p})},S(\mathbf{p}))e^{i(-\omega_{S(\mathbf{p})}t-S(\mathbf{p})\cdot\mathbf{x})}\right]
\end{split}\end{equation}
for which we introduce the shorthand notation\footnote{The choice not to include the $1/\sqrt{2\omega_\mathbf{p}}$ factor in the normalization of $a_\mathbf{p}$ and $b_{\mathbf{p}}$ is deliberate and results in much more compact results for the Noether charges.} 
\begin{equation}
\phi(x)=\int\frac{d^3p}{2\omega_\mathbf{p}\sqrt{p_4/\kappa}}\left[a_{\mathbf{p}}e^{ipx}+b^\dagger_{\mathbf{p}}e^{iS(p)x}\right]
\label{eq-mode-phi}
\end{equation}
where we define the positive and negative frequency mode coefficients
\begin{equation}
a_\mathbf{p}=\sqrt{\frac{\kappa}{p_4}}\tilde{\phi}(\omega_\mathbf{p},\mathbf{p}),\quad b_{\mathbf{p}}=\sqrt{\frac{\kappa}{p_4}}\frac{\kappa^3}{p_+^3}\tilde{\phi}^*(-\omega_{S(\mathbf{p})},S(\mathbf{p})).
\end{equation}
Since, after quantization, one expects the coefficients to correspond to creation/annihilation operators for particles and antiparticles, we will refer to the modes in the following briefly as particle and antiparticle modes.

Accordingly, the mode expansion of the conjugate field is
\begin{equation}
\phi^\dagger(x)=\int\frac{d^3p}{2\omega_\mathbf{p}\sqrt{p_4/\kappa}}\left[a^\dagger_{\mathbf{p}}e^{iS(p)x}+b_{\mathbf{p}}e^{ipx}\right].
\label{eq-mode-phid}
\end{equation}
An action for the field, invariant under $\kappa$-Poincaré symmetries, can be written as 
\begin{equation}
\mathcal{S}(\phi^\dagger,\phi)=-\int d^4 x\left[S(\partial_\mu)\phi^\dagger\star\partial^\mu\phi+m^2\phi^\dagger\star\phi\right]\equiv\mathcal{S}_1
\label{eq-s1}
\end{equation}
with the Lagrangian
\begin{align}
\mathcal{L}(\phi^\dagger,\phi)=-S(\partial_\mu)\phi^\dagger\star\partial^\mu\phi-m^2\phi^\dagger\star\phi\equiv\mathcal{L}_1,
\label{eq-l1}
\end{align}
where the $\star$-product is defined via the Weyl map \eqref{embeddingWeylMap} as
\begin{equation}
f\star g=\mathcal{W}^{-1}\left(\mathcal{W}(f)\mathcal{W}(g)\right)\ .
\end{equation}
The action~(\ref{eq-s1}) is the same, through the Weyl map ($\hat{{\cal S}} (\hat{\phi}^{\dagger},\hat{\phi}) = {\cal W}(\mathcal{S}(\phi^\dagger,\phi))$), as the action on noncommutative fields
\begin{equation}
\hat{{\cal S}}\left(\hat{\phi}^{\dagger},\hat{\phi}\right)
= - \hat{\int} \left[ \hat{\partial}^\dagger_\mu \hat{\phi}^\dagger \hat{\partial}^\mu\hat{\phi} + m^2 \hat{\phi}^\dagger \hat{\phi}\right] \ ,
\label{eq-S1-noncomm}
\end{equation}
where the noncommutative derivatives are defined in App.~\ref{app:calculus}.

\section{Noether analysis}
\label{sec:complex noether}
In this section we present the derivation of Noether charges associated with $\kappa$-Poincaré symmetries for the complex scalar field described by \eqref{eq-l1}. 
We then compute the symplectic structure associated with the action, and show that the charges satisfy the Poincar\'e algebra, so that they are indeed a representation of the (deformed) symmetries of the theory that preserve Lorentz (and Poincar\'e) invariance.
The analysis is based on techniques that were used in particular in~\cite{FreJurekNowkfield2} and~\cite{kDiscrete1}, and we refer to those papers for further details. We review the relevant properties of the differential calculus in Apppendix ~\ref{app:calculus}, as well as some other useful formulas in Appendices~\ref{app:plane waves}, \ref{app:star x} and~\ref{app:AntipodeMeasure}.
The Lorentz charges are here calculated explicitly, with the canonical approach.

As it was shown in~\cite{FreJurekNowkfield2} (see also~\cite{kDiscrete1}), the 5D differential calculus can be described in terms of the sum of two differentials, one associated with a 5D translation and one with Lorentz transformations: $d=d_T + d_L$ (we are going to define them explicitly below). 
It should be noted that the differentials associated with translations ($d_T$) and Lorentz transformations ($d_L$) do satisfy Leibniz rule under the $\star$-product, 
\begin{equation}\begin{split}
\label{eq-leibniz}
d(f\star g)=d f\star g+f\star d g
\end{split}\end{equation}
which is a crucial property in the calculations that follow. This is made possible by having both the derivative operators and parameters (that is, both $\partial_A$ and $dx^A$ in the case of $d_T$) noncommutative.

Moreover, the derivatives emerging from the differential $d_T$ generate translations that close the standard Poincaré algebra with the Lorentz generators, as shown in detail in~\cite{FreJurekNowkfield2}.
This means that with this definition of translations, one expects to find Noether charges that satisfy the standard Poincaré algebra, i.e. that close the so-called classical basis of $\kappa$-Poincaré~\cite{LukKosMasClassicalBasis}.
In this particular basis, the algebra is undeformed, all the deformation being in the coalgebra structures (see App.~\ref{app:kPoincare}).

\subsection{Variation of the action}\label{sec:variation-l1}
Following the properties of integration by parts with respect to the $\star$-product here adopted (see ~\cite{FreJurekNowkfield2,kDiscrete1} and App.~\ref{app:calculus}), the variation of the action \eqref{eq-s1} can be expressed as 
\begin{equation}
\begin{split}
\delta \mathcal{S}_1=\int d^{4}x\left[\partial_{A}\left(\Pi_{1}^{A}\phi^{\dagger}\star\delta\phi\right)+S(\partial_{A})\left(\delta\phi^{\dagger}\star S(\Pi_{1}^{A})\phi\right)\right.\\
\left.-\left(S(\partial_\mu)S(\partial^\mu)+m^2\right)\phi^\dagger\star\delta\phi-\delta\phi^\dagger\star  \left(\partial_\mu\partial^\mu+m^2\right)\phi\right],
\end{split}\label{eq-vars1}\end{equation}
where the $\Pi^A_1$ operators denote
\begin{equation}
\begin{gathered}
\Pi_{1}^{0}=\frac{p_4}{\kappa} \partial^{0} \ , \quad
\Pi_{1}^{j} =\frac{p_4}{\kappa}\partial^{j} \ , \quad 
\Pi_{1}^{4} =-i\frac{m^{2}}{\kappa} \ .
\end{gathered}
\end{equation}
Ignoring the surface terms, we read off the equations of motion, which,
since the mass Casimir\footnote{We refer to the Laplacian $\partial_\mu \partial^\mu$ as the mass Casimir, since it corresponds to the quadratic invariant of the Poincar\'e algebra.} is invariant with respect to the antipode, are undeformed for both $\phi$ and $\phi^\dagger$:
\begin{gather}
\left(\partial_\mu\partial^\mu+m^2\right)\phi =0\label{eq-eom-1}\ ,\\
\left(S(\partial_\mu)S(\partial^\mu)+m^2\right)\phi^\dagger=\left(\partial_\mu\partial^\mu+m^2\right)\phi^\dagger =0\ .
\label{eq-eom-2}
\end{gather}
The basis for our calculations of the Noether charges is then the variation of the Lagrangian. Imposing \eqref{eq-eom-1} and \eqref{eq-eom-2}, it becomes
\begin{equation}
\begin{split}
\delta\mathcal{L}_1=\partial_A\left(\Pi^A_1\phi^\dagger\star\delta\phi\right)+S(\partial_A)\left(\delta\phi^\dagger\star S(\Pi^A_1)\phi\right) \ .
\end{split}
\label{eq-current-base}
\end{equation}

\subsection{Translation charge}

The 5D translation differential takes the form~\cite{SitarzDiffCalc,FreJurekNowkfield2,kDiscrete1} $d_T=\epsilon^A\partial_A=dx^\mu\star\partial_\mu+dx^4\star\partial_4$ where the $\star$-commutators of the $dx^A$ are reported in App.~\ref{app:noncomm parameters}.
To calculate the translation charge, we substitute the differential $d_T$ in place of the variation in \eqref{eq-current-base} which we can do because $d_T$ satisfies the Leibniz rule):
\begin{equation}
\label{eq-varl1-t}
d_T\mathcal{L}_1-\partial_{A}\left(\Pi_{1}^{A}\phi^{\dagger}\star d_T\phi\right)-S(\partial_{A})\left(d_T\phi^{\dagger}\star S(\Pi_{1}^{A})\phi\right)=0
\end{equation}
Since $\epsilon^B$ is itself noncommutative, in order to factor it out we commute it to the left in all terms in \eqref{eq-varl1-t} using \eqref{eq-leibniz}:
\begin{equation}
\begin{split}
\epsilon^{B}\partial_{B}\left[\mathcal{L}_{1}-\partial_{A}\left(\Pi_{1}^{A}\phi^{\dagger}\star\phi\right)\right]+\partial_{A}\left(\epsilon^{B}\partial_{B}\Pi_{1}^{A}\phi^{\dagger}\star\phi\right)\\-S(\partial_{A})\left(\epsilon^{B}\partial_{B}\phi^{\dagger}\star S(\Pi_{1}^{A})\phi\right)=0
\end{split}
\label{eq-varl1-t-1}
\end{equation}
The first term vanishes on-shell (as evident from \eqref{eq-vars1}), and since~\cite{FreJurekNowkfield2} $\partial_A\epsilon^B=0$, we obtain the energy-momentum conservation equation\footnote{The sign change with respect to \eqref{eq-varl1-t-1} in the definition of $T^A_{\;B}$ comes from our choice to define the positive energy modes as $e^{ipx}$ rather than $e^{-ipx}$.} 
\begin{equation}
\partial_{A}\left(\partial_{B}\Pi_{1}^{A}\phi^{\dagger}\star\phi\right)-S(\partial_{A})\left(\partial_{B}\phi^{\dagger}\star S(\Pi_{1}^{A})\phi\right)\equiv-\partial_A T^A_{\;B}=0
\end{equation}
With the explicit forms \eqref{eq-sd0}-\eqref{eq-sd4} of $S(\partial_A)$, we find that the time component of the current is
\begin{equation}
T^0_{\;B}=-\partial_B\Pi^0_1\phi^\dagger\star\phi-\partial_B\phi^\dagger\star S(\Pi^0_1)\phi
\end{equation}
Integrating it over space and substituting the mode expansions \eqref{eq-mode-phi} and \eqref{eq-mode-phid}, we obtain the translation charges
\begin{equation}
\begin{split}
\mathcal{P}^1_\mu=&\;\int d^3x\;T^0_{\;\mu}\\
=&\;\int d^3 x\left[\left(\partial_{\mu}\Pi_{1}^{0}\phi^{\dagger}\star\phi\right)+\left(\partial_{\mu}\phi^{\dagger}\star S(\Pi_{1}^{0})\phi\right)\right]\\
=&\;\int d^{3}x\frac{d^{3}p}{2\omega_{\mathbf{p}}}\frac{d^{3}q}{2\omega_{\mathbf{q}}}\frac{\kappa}{p_{4}}\left[\left(S(p_{\mu})\frac{p_{4}}{\kappa}S(\omega_{\mathbf{p}})a_{\mathbf{p}}^{\dagger}e^{iS(p)x}\right.\right.\\
&+\left.p_{\mu}\frac{p_{4}}{\kappa}\omega_{\mathbf{p}}b_{\mathbf{p}}e^{ipx}\right)\star\left(a_{\mathbf{q}}e^{iqx}+b^\dagger_{\mathbf{q}}e^{iS(q)x}\right)\\
&+\left(S(p_{\mu})a_{\mathbf{p}}^{\dagger}e^{iS(p)x}+p_{\mu}b_{\mathbf{p}}e^{ipx}\right)\\
&\star\left.\left(\frac{p_{4}}{\kappa}S(\omega_{\mathbf{q}})a_{\mathbf{q}}e^{iqx}+\frac{p_{4}}{\kappa}\omega_{\mathbf{q}}b^\dagger_{\mathbf{q}}e^{iS(q)x}\right)\right]\\
=&\;\int \frac{d^{3}p}{2\omega_\mathbf{p}}\left[\frac{p_+^3}{\kappa^3}b_{\mathbf{p}}^\dagger b_{\mathbf{p}}p_{\mu}-a_{\mathbf{p}}^\dagger a_{\mathbf{p}}S(p_{\mu})\right]
\end{split}
\label{eq-pmu-l1}
\end{equation}

\begin{equation}
\mathcal{P}^1_4=\int d^3x\;T^0_{\;4}=\int \frac{d^{3}p}{2\omega_\mathbf{p}}\left[\frac{p_+^{3}}{\kappa^{3}}b_{\mathbf{p}}^\dagger b_{\mathbf{p}}+a_{\mathbf{p}}^\dagger a_{\mathbf{p}}\right](p_4-\kappa)
\label{eq-p4-l1}
\end{equation}
where we evaluated the Dirac deltas according to the prescriptions in Appendix~\ref{app:plane waves}.

\subsection{Lorentz charges}
\label{sec:Lorentz charges}
It has been shown in~\cite{FreJurekNowkfield2} (see also~\cite{kDiscrete1}) that the differential associated with Lorentz transformations that is compatible with the differential calculus is $d_L=\omega^{\mu\nu} \star \left(x_\mu\star\frac{\kappa}{\Delta_+}\partial_\nu\right)\equiv\omega^{\mu\nu} \star L_{\mu\nu}$ where $\omega^{\mu\nu}=-\omega^{\nu\mu}$, $\partial_A\omega^{\mu\nu}=0$ and $d_L$ satisfies the Leibniz rule \eqref{eq-leibniz} (the $\star$-commutation properties of $\omega^{\mu\nu}$ are discussed in App.~\ref{app:noncomm parameters}). Replacing the variation in \eqref{eq-current-base} with this differential (which we can do because $d_L$ also satisfies the Leibniz rule), we proceed as we did for translations. Since the operator $L_{\mu\nu}$ does not commute with $\partial_A$, commuting $\omega^{\mu\nu}$ to the left requires an extra step:
\begin{equation}\begin{split}
L_{\mu\nu}\mathcal{L}_1+[\partial_{A},L_{\mu\nu}](\Pi_{1}^{A}\phi^\dagger\star\phi)+L_{\mu\nu}\partial_{A}(\Pi_{1}^{A}\phi^\dagger\star\phi)\\-\partial_{A}(L_{\mu\nu}\Pi_{1}^{A}\phi^\dagger\star\phi)+S(\partial_{A})(L_{\mu\nu}\phi^\dagger\star S(\Pi_{1}^{A})\phi)=0
\end{split}\end{equation}
A straightforward calculation gives the explicit commutator $[\partial_A,L_{\mu\nu}]=\eta_{A[\mu}\partial_{\nu]}$, which, together with the explicit form of $S(\partial_A)$, results in the conservation equation for Lorentz transformations $\partial_A\mathcal{J}^A_{\mu\nu}=0$ with
\begin{equation}
\begin{split}
\mathcal{J}^0_{\mu\nu} & =  x_{[\mu}\star \left( \frac{\kappa}{\Delta_+}\partial_{\nu]}\Pi_{1}^{0}\phi^\dagger\star\phi + \frac{\kappa}{\Delta_+}\partial_{\nu]}\phi^\dagger\star S(\Pi_{1}^{0})\phi \right) \\ & ~~~~ +\eta_{A[\nu}\delta_{\mu]0}\Pi_{1}^{A}\phi^\dagger\star\phi \ .
\label{eq-lorentz-current}
\end{split}
\end{equation}
The calculation then proceeds much like for translations in \eqref{eq-pmu-l1}, with a few important caveats. Firstly, we make use of the following identity:
\begin{align}
x_\mu\star\phi(x)=\frac{1}{\kappa}\left(x_\mu\Delta_++ix_0\partial_\mu\right)\phi(x)
\label{eq-x-star}
\end{align}
proven in Appendix~\ref{app:star x}. Secondly, special care must be taken with the Fourier transform of $x_j$. Due to the deformed addition rules for plane waves (see Appendix~\ref{app:plane waves}):
\begin{equation}\begin{split}
e^{ipx}\star e^{iqx}=e^{i(p\oplus q)_\mu x^\mu}=e^{i(p\oplus q)_0 x^0}e^{i\left(\frac{q_+}{\kappa}p_j+q_j\right)x^j}
\end{split}\end{equation}
$x_j$ must be Fourier transformed (schematically) as
\begin{equation}\begin{split}
x_j f(x)\star g(x)\rightarrow \tilde f(p)\tilde g(q)e^{i(p\oplus q)_0 x^0}\left(-i\frac{\kappa}{q_+}\frac{\partial}{\partial p^j}\right)e^{i(p\oplus q)_j x^j}\label{eq-fourier-x}
\end{split}\end{equation}
To better illustrate this methodology, we present the derivation of the antiparticle ($b^\dagger_{\mathbf p},b_{\mathbf{p}}$) sector $\mathcal{N}_j^{bb}$ for the boost charge:
\begin{equation}
\begin{split}
\mathcal{N}_j^{bb}=&\,\int d^{3}x\frac{d^{3}p}{2\omega_{\mathbf{p}}}\frac{d^{3}q}{2\omega_{\mathbf{q}}}\frac{\kappa}{p_4}\left[\frac{\kappa}{p_{+}}x_{[0}\star\left(\partial_{j]}\Pi_{1}^{0}b_{\mathbf{p}}e^{ipx}\star b_{\mathbf{q}}^\dagger e^{iS(q)x}\right.\right.\\
&\left.+\partial_{j]}b_{\mathbf{p}}e^{ipx}\star S(\Pi_{1}^{0})b_{\mathbf{q}}^\dagger e^{iS(q)x}\right)\\
&\left.-\eta_{jA}\Pi_{1}^{A}b_{\mathbf{p}}e^{ipx}\star b_{\mathbf{q}}e^{iS(q)x}\right]\\
=&\,\int d^{3}x\frac{d^{3}p}{2\omega_{\mathbf{p}}}\frac{d^{3}q}{2\omega_{\mathbf{q}}}\left[-\frac{\kappa}{p_{+}}x_{[0}p_{j]}\star \left(\omega_{\mathbf{p}}+\omega_{\mathbf{q}}\right)+ip_{j}\right]\\
&\times b_{\mathbf{p}}b_{\mathbf{q}}^{\dagger}e^{i(p\oplus S(q))x}
\end{split}
\end{equation}
At this point we use \eqref{eq-x-star} to obtain
\begin{equation}
\begin{split}
\frac{\kappa}{p_{+}}x_{[0}p_{j]}\star e^{i(p\oplus S(q))x}=&\,\left(\frac{\kappa}{q_{+}}x_{[0}-\frac{1}{p_{+}}x_{0}(p\oplus S(q))_{[0}\right)\\
&\times p_{j]}e^{i(p\oplus S(q))}
\end{split}
\end{equation}
Note that after evaluating the Dirac delta according to \eqref{eq-delta3-ps} the second term vanishes, since $\mathbf{p}\oplus S(\mathbf{p})=0$ and $\omega_{\mathbf{p}}\oplus S(\omega_{\mathbf{p}})=0$ by definition. We are thus left with
\begin{align}
\begin{split}
\mathcal{N}_{j}^{bb}=&\,\int d^{3}x\frac{d^{3}p}{2\omega_{\mathbf{p}}}\frac{d^{3}q}{2\omega_{\mathbf{q}}}\left[-\frac{\kappa}{q_{+}}\left(tp_{j}-x_{j}\omega_{\mathbf{p}}\right)\left(\omega_{\mathbf{p}}+\omega_{\mathbf{q}}\right)+ip_{j}\right]\\
&\times b_{\mathbf{p}}b_{\mathbf{q}}^{\dagger}e^{i(p\oplus S(q))x}
\end{split}
\end{align}
The terms proportional to $tp_j$ and $p_j$ can be simplified by resolving the exponents as Dirac deltas to
\begin{equation}
\int\frac{d^{3}p}{2\omega_{\mathbf{p}}}\frac{p_{+}^{3}}{\kappa^{3}}\left[-\frac{\kappa}{p_{+}}tp_j+\frac{i}{2}\frac{p_{j}}{\omega_\mathbf{p}}\right]b_{\mathbf{p}}b_{\mathbf{p}}^{\dagger}
\label{eq-njbb-1}
\end{equation}
The remaining terms are evaluated using \eqref{eq-fourier-x}:
\begin{equation}
\begin{split}
&\frac{1}{2}\int d^{3}xd^{3}p\frac{d^{3}q}{2\omega_{\mathbf{q}}}\frac{\kappa}{q_{+}}x_{j}\left(\omega_{\mathbf{p}}+\omega_{\mathbf{q}}\right)b_{\mathbf{p}}b_{\mathbf{q}}^{\dagger}e^{i(p\oplus S(q))x}\\
=&\,\frac{1}{2}\int d^{3}xd^{3}p\frac{d^{3}q}{2\omega_{\mathbf{q}}}\frac{\kappa}{q_{+}}\left(\omega_{\mathbf{p}}+\omega_{\mathbf{q}}\right)b_{\mathbf{p}}b_{\mathbf{q}}^{\dagger}e^{i(p\oplus S(q))_{0}x^{0}}\\
&\times\left(-i\frac{q_{+}}{\kappa}\frac{\partial}{\partial p^{j}}\right)e^{i\frac{\kappa}{q_{+}}\left(p-q\right)_{i}x^{i}}\\
=&\,\frac{i}{2}\int d^{3}xd^{3}p\frac{d^{3}q}{2\omega_{\mathbf{q}}}\left[\frac{\partial}{\partial p^{j}}b_{\mathbf{p}}\left(\omega_{\mathbf{p}}+\omega_{\mathbf{q}}\right)e^{i(p\oplus S(q))_{0}x^{0}}\right]\\
&\times b_{\mathbf{q}}^{\dagger}e^{i(p\oplus S(q))_{i}x^{i}}\\
=&\,\int\frac{d^{3}p}{2\omega_{\mathbf{p}}}\frac{p_+^3}{\kappa^3}\left[i\omega_{\mathbf{p}}\frac{\partial b_{\mathbf{p}}}{\partial p^{j}}b_{\mathbf{p}}^{\dagger}-\frac{i}{2}\frac{p_{j}}{\omega_{\mathbf{p}}}b_{\mathbf{p}}b_{\mathbf{p}}^{\dagger}+\frac{\kappa}{p_{+}}\frac{p_{j}}{\omega_{\mathbf{p}}}b_{\mathbf{p}}b_{\mathbf{p}}^{\dagger}\right]
\end{split}
\label{eq-njbb-2}
\end{equation}
Collecting \eqref{eq-njbb-1} and \eqref{eq-njbb-2}, we obtain
\begin{equation}
\mathcal{N}_j^{bb}=i\int\frac{d^{3}p}{2\omega_{\mathbf{p}}}\frac{p_+^3}{\kappa^3}\omega_{\mathbf{p}}b_{\mathbf{p}}^{\dagger}\frac{\partial b_{\mathbf{p}}}{\partial p^{j}}.
\end{equation}
Applying the same method to the remaining contributions to $\mathcal{J}^0_{0j}$ (and similarly for rotations $\mathcal{J}^0_{ij}$), we obtain the rotation ($\mathcal{M}_j^1$) and boost ($\mathcal{N}_j^1$) charges, which in the end take a fairly simple form:
\begin{equation}\begin{split}
\label{rotation charge}
\mathcal{M}^1_{k}&=i\epsilon_{ijk}\int\frac{d^{3}p}{2\omega_\mathbf{p}}\left[\frac{p_{+}^{3}}{\kappa^{3}}p_ib^\dagger_{\mathbf{p}}\frac{\partial b_{\mathbf{p}}}{\partial p^j}+S(p_i)a^\dagger_\mathbf{p}\frac{\partial a_\mathbf{p}}{\partial S(p^j)}\right]
\end{split}\end{equation}

\begin{equation}\begin{split}
\label{boost charge}
\mathcal{N}^1_{j}&=i\int\frac{d^{3}p}{2\omega_\mathbf{p}}\left[\frac{p_{+}^{3}}{\kappa^{3}}b^\dagger_{\mathbf{p}}\frac{\partial b_{\mathbf{p}}}{\partial p^j}\omega_\mathbf{p}+a^\dagger_\mathbf{p}\frac{\partial a_\mathbf{p}}{\partial S(p^j)}S(\omega_\mathbf{p})\right]
\end{split}\end{equation}

The Lorentz charges~(\ref{rotation charge}) and~(\ref{boost charge}) are here obtained for the first time explicitly in a direct, canonical approach\footnote{Some parts of the derivation were already in~\cite{FreJurekNowkfield2}. However a complete explicit derivation of the charges was missing.}. Their expression must be compared with the one derived in~\cite{JurekAndreaDiscrete} through a different technique, based on a ``covariant phase space'' approach, which did not directly depend on the surface terms in the deformed action. We will discuss the comparison in section~\ref{sec:comparison}.

\subsection{Symplectic structure and algebra}
The starting point for obtaining the symplectic form is the symplectic potential current $\theta^{\mu}$, which comes from the variation of a Lagrangian in standard field theory as
\begin{align}
\delta\mathcal{L}(\phi)=E.O.M\times\delta\phi+\partial_\mu\theta^\mu
\end{align}
This definition naturally extends to our deformed case, and we identify $\theta^A$ as the total derivative term in
\eqref{eq-vars1}, whose time component is
\begin{equation}\begin{split}
\theta^0=\Pi^0_1\phi^\dagger\star\delta\phi-\delta\phi^\dagger\star S(\Pi^0_1)\phi
\end{split}\end{equation}
The symplectic form $\Omega$ is obtained by integrating the antisymmetrized second variation of $\theta^0$ over a constant time Cauchy hypersurface $\Sigma_t$:
\begin{equation}\label{symplectic_form}
\begin{split}
\Omega&=\int_{\Sigma_t} d^3 x\left[\Pi^0_1\delta\phi^\dagger\wedge_{\star}\delta\phi-\delta\phi^\dagger\wedge_{\star} S(\Pi^0_1)\delta\phi\right]\\
&=i\int \frac{d^3p}{2\omega_\mathbf{p}}\left[\frac{p_+^3}{\kappa^3}\delta b_{\mathbf{p}}\wedge\delta b^\dagger_{\mathbf{p}}-\delta a^\dagger_{\mathbf{p}}\wedge\delta a_{\mathbf{p}}\right],
\end{split}\end{equation}
where we denoted $a \wedge_{\star}b=a\star b-b\star a$.
The resulting Poisson brackets are
\begin{align}
\{a_\mathbf{p},a^\dagger_\mathbf{q}\}&=-2i\omega_\mathbf{p}\delta^3(\mathbf{p}-\mathbf{q})\label{eq-poisson-a}\\
\{b_{\mathbf{p}},b^\dagger_{\mathbf{p}}\}&=-2i\omega_\mathbf{p}\frac{\kappa^3}{p_+^3}\delta^3(\mathbf{p}-\mathbf{q})\label{eq-poisson-b}
\end{align}

It can be shown that with the brackets \eqref{eq-poisson-a} and \eqref{eq-poisson-b} $\mathcal{P}_\mu^1$, $\mathcal{M}_j^1$ and $\mathcal{N}_j^1$ satisfy the standard Poincar\'e algebra (as expected for the classical basis of $\kappa$-Poincar\'e),  assuring the relativistic invariance of the theory:

\begin{equation}
\begin{split}
\{\mathcal{P}_\mu,\mathcal{P}_\nu\}=0,\quad\{\mathcal{N}_j,\mathcal{P}_k\}=\eta_{jk}\mathcal{P}_0,\\\quad\{\mathcal{N}_j,\mathcal{P}_0\}=-\mathcal{P}_j,\quad
\{\mathcal{M}_j,\mathcal{P}_k\}=\epsilon_{jkl}\mathcal{P}_l,\\\quad\{\mathcal{M}_j,\mathcal{P}_0\}=0,\quad\{\mathcal{M}_j,\mathcal{M}_k\}=\epsilon_{jkl}\mathcal{M}_l,\\
\{\mathcal{M}_j,\mathcal{N}_k\}=\epsilon_{jkl}\mathcal{N}_l,\quad\{\mathcal{N}_j,\mathcal{N}_k\}=-\epsilon_{jkl}\mathcal{M}_l
\end{split}
\end{equation}
The proof is fully analogous to the undeformed case, since in the $b$ particle sector the extra $p_+^3/\kappa^3$ factor is canceled by the bracket \eqref{eq-poisson-b}, while the $a$ particle sector is identical  to the undeformed case with $\mathbf{p}$ changed to $-S(\mathbf{p})$.
The algebra also has 3 additional brackets
\begin{equation}\begin{split}
\left\{\mathcal{P}_\mu,\mathcal{P}_4\right\}=\left\{\mathcal{M}_j,\mathcal{P}_4\right\}=\left\{\mathcal{N}_j,\mathcal{P}_4\right\}=0
\end{split}\end{equation}
which shows that $P_4$ is a central charge for the $\kappa$-Poincar\'e algebra in the classical basis.
The fact that the charges satisfy the Poincar\'e algebra shows that they are a representation of the relativistic symmetries that preserve Lorentz (and Poincar\'e) invariance.
\label{sec:complex symplectic}

\section{Charge conjugation}
\label{sec:chargeConjugation}

\subsection{Definition}

Let us start by noticing that the definition of the discrete symmetries in a deformed theory is not free from ambiguity. Several different assumptions can be made, as long as they reduce to the standard properties of Poincaré symmetries in the undeformed (commutative) limit.

A definition of charge conjugation ${\cal C}$ that appears to be natural even in the deformed setting is to assume that its action is to switch between particle and antiparticle modes:
\begin{equation}\label{eq-c-momentum}
a_{{\bf p}}\stackrel{{\cal C}}{\rightarrow}b_{\bf p}\ , \qquad b_{\bf p}\stackrel{{\cal C}}{\rightarrow}a_{\bf p} \ .
\end{equation}
This is indeed the choice that was made in the previous works~\cite{kDiscrete1,JurekAndreaDiscrete}.
Notice that in terms of momentum-space fields this means that
\begin{equation}
\tilde{\phi}\left(\omega_{{\bf p}},{\bf p}\right)\stackrel{{\cal C}}{\rightarrow}\frac{\kappa^{3}}{p_{+}^{3}}\tilde{\phi}^{*}\left(-\omega_{S\left({\bf p}\right)},S\left({\bf p}\right)\right) \ .
\end{equation}

Using the definition~(\ref{on-shell field}) or~(\ref{eq-mode-phi}) of the field it is straightforward to show that this implies that $\cal{C}$ acts as\footnote{This action extends to noncommutative fields as $\hat{\phi}(\hat{x}) \xrightarrow[]{\mathcal{C}}\hat{\phi}^\dagger(\hat{x})$.}
\begin{equation}
\phi(x) \xrightarrow[]{\mathcal{C}}\phi^\dagger(x) \ .
\label{eq-c-def}
\end{equation}
This rule of action is the same as in the undeformed (commutative) case, except that here one should be mindful of the fact that the
\mbox{$\left(\right)^{\dagger}$} is not just complex conjugation
\mbox{$\left(\right)^{*}$}, as it also involves a transposition
of the noncommutative plane wave.

It should be noted that the action $\mathcal{S}(\phi^\dagger,\phi)$ defined in \eqref{eq-s1} is not invariant under the $\mathcal{C}$ transformation here defined, as evident from the form of its conserved charges (e.g. \eqref{eq-pmu-l1}). Indeed, this was not the action considered in~\cite{kDiscrete1,JurekAndreaDiscrete}, whose purpose was to start from a $\cal C$-invariant model. We elaborate on this point in the next two sections.

\subsection{Aside on alternative ordering in the action}
\label{sec:s2}

The ordering between $\phi$ and $\phi^\dagger$ in the action \eqref{eq-s1} is arbitrary, and one may just as well consider the action
\begin{equation}
\mathcal{S}(\phi,\phi^\dagger)=-\int d^4 x\left[S(\partial_\mu)\phi\star\partial^\mu\phi^\dagger+m^2\phi\star\phi^\dagger\right]\ ,
\label{eq-s2}
\end{equation}
which is related to $\mathcal{S}(\phi^\dagger,\phi)$ by the transformation $\mathcal{C}$ defined in \eqref{eq-c-def}.
Repeating the analysis in sec. \ref{sec:complex noether} for $\mathcal{S}(\phi,\phi^\dagger)$, we obtain the following conserved charges:
\begin{align}
\mathcal{P}^2_\mu&=\int \frac{d^{3}p}{2\omega_\mathbf{p}}\left[\frac{p_+^{3}}{\kappa^{3}}a_{\mathbf{p}}^\dagger a_{\mathbf{p}}p_{\mu}-b_{\mathbf{p}}^\dagger b_{\mathbf{p}}S(p_{\mu})\right]\label{eq-pmu-l2}\ ,\\
\mathcal{P}^2_4&=\int \frac{d^{3}p}{2\omega_\mathbf{p}}\left[\frac{p_+^{3}}{\kappa^{3}}a_{\mathbf{p}}^\dagger a_{\mathbf{p}}+b_{\mathbf{p}}^\dagger b_{\mathbf{p}}\right](p_4-\kappa) \ ,
\label{eq-p4-l2}
\end{align}
\begin{equation}\begin{split}
\mathcal{M}^2_{k}&=i\epsilon_{ijk}\int\frac{d^{3}p}{2\omega_\mathbf{p}}\left[\frac{p_{+}^{3}}{\kappa^{3}}p_ia^\dagger_\mathbf{p}\frac{\partial a_\mathbf{p}}{\partial p^j}+S(p_i)b^\dagger_{\mathbf{p}}\frac{\partial b_{\mathbf{p}}}{\partial S(p^j)}\right] \ ,
\label{rotation charge S2}
\end{split}\end{equation}
\begin{equation}\begin{split}
\mathcal{N}^2_{j}&=i\int\frac{d^{3}p}{2\omega_\mathbf{p}}\left[\frac{p_{+}^{3}}{\kappa^{3}}a^\dagger_\mathbf{p}\frac{\partial a_\mathbf{p}}{\partial p^j}\omega_\mathbf{p}+b^\dagger_{\mathbf{p}}\frac{\partial b_{\mathbf{p}}}{\partial S(p^j)}S(\omega_\mathbf{p})\right] \ .
\label{boost charge S2}
\end{split}\end{equation}
It is immediately apparent that these charges are also related to the ones obtained from $\mathcal{S}(\phi^\dagger,\phi)$ by the $\mathcal{C}$ transformation~\eqref{eq-c-momentum}. This shows that the ordering of $\phi$ and $\phi^\dagger$ is responsible for determining which particle species ends up in the $p$ and $S(p)$ momentum spaces (cf. Fig. \ref{fig:-momentum-space}).

This fact can be explained by a closer examination of the steps leading up to the calculation of~\eqref{eq-pmu-l1}. Since the translation parameter $\epsilon^B$ is always commuted to the left, in the end $\partial_B$ always acts on the left field, which for particles corresponds schematically to $\partial_B a^\dagger_\mathbf{p}e^{iS(p)x}\star a_\mathbf{p}e^{ipx}$ for $\mathcal{S}(\phi^\dagger,\phi)$ and $\partial_B a_\mathbf{p}e^{ipx}\star a^\dagger_\mathbf{p}e^{iS(p)x}$ for $\mathcal{S}(\phi,\phi^\dagger)$.
This crucial point, regarding the commutation of the parameter in the Lagrangian, and determining the difference between the charges (\ref{eq-p4-l2}),(\ref{rotation charge S2}),(\ref{boost charge S2})  and (\ref{eq-pmu-l1}),(\ref{rotation charge}),(\ref{boost charge}), can be uncovered only through the canonical formalism, and it was impossible to  appreciate through the covariant phase space approach. Indeed, it was overlooked in the previous works, obscuring the cause of the breakdown of the $\cal C$ invariance of the charges calculated there. We will discuss the consequences of this observation in Sec.~\ref{sec:comparison}.
It is important for the following discussion to point out that the difference in commuting the parameter to the left or to the right in the Lagrangian amounts to a total derivative acting on the fields.

One should note, in closing this section, that the theories described by $\mathcal{S}(\phi^\dagger,\phi)$ and $\mathcal{S}(\phi,\phi^\dagger)$ can be considered physically equivalent. While they assign different momentum properties to particles and antiparticles, the only way to distinguish them would be by those very properties; thus, the labeling itself is irrelevant.
On the other hand, once the labeling convention is fixed, the two theories become distinct (as they assign different momenta to the same species of particle) and related by $\mathcal{C}$, and as such neither of them is $\mathcal{C}$-invariant.

\subsection{Incompatibility between $\mathcal{C}$ and Poincaré invariance}
\label{sec:symmetric action}
A natural approach to compensating for the loss of charge conjugation symmetry is to consider the symmetrized Lagrangian
\begin{equation}
\mathcal{L}_C=\frac{1}{2}\left(\mathcal{L}(\phi^\dagger,\phi)+\mathcal{L}(\phi,\phi^\dagger)\right) \ .
\label{eq-lc}
\end{equation}
This was the starting point of the studies carried out in~\cite{kDiscrete1,JurekAndreaDiscrete}.
The Noether analysis, performed through a direct computation using the canonical formalism as in sec.~\ref{sec:complex noether}, results in the sum of the charges obtained for $\mathcal{L}(\phi^\dagger,\phi)$ and $\mathcal{L}(\phi,\phi^\dagger)$, e.g. for translations
\begin{equation}
\begin{split}
\mathcal{P}_\mu^C&=\frac{1}{2}\left(\mathcal{P}_\mu^1+\mathcal{P}_\mu^2\right)\\
&=\frac{1}{2}\int\frac{d^3p}{2\omega_\mathbf{p}}\left[\frac{p_+^3}{\kappa^3}p_\mu-S(p_\mu)\right]\left(a^\dagger_\mathbf{p} a_\mathbf{p}+b^\dagger_{\mathbf p}b_{\mathbf{p}}\right)
\end{split}
\end{equation}
The same happens for the Lorentz charges $\mathcal{M}_j^C$ and $\mathcal{N}_j^C$ for the action~(\ref{eq-lc}), which turn out just to be the sums of the charges~(\ref{rotation charge}) and (\ref{boost charge}) with (\ref{rotation charge S2}) and (\ref{boost charge S2}) respectively.
All these resulting charges differ with respect to the ones calculated in~\cite{kDiscrete1,JurekAndreaDiscrete}. In particular, differently from those in~\cite{kDiscrete1,JurekAndreaDiscrete}, $\mathcal{P}_\mu$, $\mathcal{M}_j^C$ and $\mathcal{N}_j^C$ are manifestly ${\cal C}$-invariant.
The reason for this discrepancy will become clearer with the following observations.

First, one can easily show (it is enough to sum eq.~(\ref{symplectic_form}) with the same expression but with $a_{\bf p}$ and $b_{\bf p}$ interchanged) that the Lagrangian~(\ref{eq-lc}) gives rise to the Poisson brackets
\begin{equation}
\{a_\mathbf{p},a^\dagger_\mathbf{q}\}=\{b_{\mathbf{p}},b^\dagger_{\mathbf{q}}\}=\frac{-2i\omega_\mathbf{p}}{1+p_+^3/\kappa^3}\delta^3(\mathbf{p}-\mathbf{q})\ .
\label{eq-poisson-c}
\end{equation}
These Poisson brackets are consistent, up to a different normalization, with the ones found in~\cite{kDiscrete1} and~\cite{JurekAndreaDiscrete}.

However, if one now uses~(\ref{eq-poisson-c}) to compute the algebra of the charges $\mathcal{P}_\mu^C$, $\mathcal{M}_k^C$ and $\mathcal{N}_j^C$, one finds that the brackets do not close, in the sense that it is not possible to express them just in terms of the charges themselves, and in particular they do not reproduce the Poincar\'e algebra.
This indicates that the relativistic symmetries are broken.
This is complementary to what was obtained in~\cite{kDiscrete1,JurekAndreaDiscrete}, where the charges associated to~(\ref{eq-lc}) were Poincar\'e invariant, but broke $\cal C$ invariance (even if the starting Lagrangian~(\ref{eq-lc}) is manifestly $\cal C$ invariant).

A tentative explanation for this puzzling fact can be given in terms of the transformation properties of $p_\mu$ and $S(p_\mu)$ under $\kappa$-Poincar\'e. Since the $a_\mathbf{p}$ and $b_{\mathbf{p}}$ sectors are independent, we can interpret the algebra of charges for $\mathcal{S}(\phi^\dagger,\phi)$ as the fact that $p_\mu$ transforms as a vector under Lorentz transformations $L_{\rho\sigma}$, while $S(p_\mu)$ transforms as a vector under antipodes of Lorentz transformations $S(L_{\rho\sigma})$ in the Hopf-algebraic sense (this was already noticed in~\cite{JurekMicheleDiscrete}).

From this perspective, the symmetrized Lagrangian \eqref{eq-lc} effectively assigns $p_\mu+S(p_\mu)$ momenta to both species of particle. However, $p_\mu+S(p_\mu)$ is not a vector under $L_{\rho\sigma}+S(L_{\rho\sigma})$, thus the theory described by $\mathcal{L}_C$ is actually not Poincar\'e invariant.

A possible workaround is to take advantage of the twisted cyclicity under integration (defined in App.~\ref{app:twisted}) to commute the terms coming from $\mathcal{L}(\phi,\phi^\dagger)$ in the action:
\begin{equation}\label{action_rewriting}
\begin{split}
\mathcal{S}_C=&\;-\frac{1}{2}\int d^4x\left[S(\partial_\mu)\phi^\dagger\star\partial^\mu\phi+m^2\phi^\dagger\star\phi\right.\\
&\left.+S(\partial_\mu)\phi\star\partial^\mu\phi^\dagger+m^2\phi\star\phi^\dagger\right]\\
=&\;-\frac{1}{2}\int d^4x\left[\vphantom{\frac{\Delta_+^3}{\kappa^3}}S(\partial_\mu)\phi^\dagger\star\partial^\mu\phi+m^2\phi^\dagger\star\phi\right.\\
&\left.+\partial^\mu\phi^\dagger\star \frac{\Delta_+^3}{\kappa^3}S(\partial_\mu)\phi+m^2\phi^\dagger\star\frac{\Delta_+^3}{\kappa^3}\phi\right]
\end{split}
\end{equation}
While this restores the Poincar\'e invariance of the action (as explained below), it corresponds to a modification of the Lagrangian by a boundary term, which affects the conserved charges.
Our study shows that this corresponds to what was implicitly done by the covariant phase formalism in~\cite{kDiscrete1,JurekAndreaDiscrete}, as we are going to discuss in the next section.

Performing once again the Noether analysis, we find charges of the form
\begin{equation}\label{charges_cps}
\mathcal{P}_\mu^{C1}=\frac{1}{2}\int\frac{d^3p}{2\omega_\mathbf{p}}\left(1+\frac{p_+^3}{\kappa^3}\right)\left[b_{\mathbf{p}}^\dagger b_{\mathbf{p}}p_\mu-a_{\mathbf{p}}^\dagger a_{\mathbf{p}}S(p_\mu)\right]
\end{equation}
with the symplectic structure identical to \eqref{eq-poisson-c}. These satisfy again the Poincar\'e algebra through the same mechanism as explained in sec.~\ref{sec:complex symplectic} (as the overall factor in~\eqref{charges_cps} exactly matches the one in~\eqref{eq-poisson-c}), but are evidently not $\mathcal{C}$-invariant anymore - both of which can be traced back to the introduction of the boundary term.

This observation suggests that, within the deformed framework, $\mathcal{C}$-invariance and Poincar\'e invariance may not be simultaneously realizable, as enforcing one appears to break the other.

\subsection{Comparison with the covariant phase space approach}
\label{sec:comparison}

We discuss now briefly how the covariant phase space approach used in~\cite{kDiscrete1,JurekAndreaDiscrete} works, highlighting some its underlying assumptions.
The idea at its basis is that using the symplectic form \eqref{symplectic_form}, one can compute all the charges using the defining relation
\begin{equation}\label{cps_approach}
    \delta_\xi \lrcorner \, \Omega = -\delta Q_\xi
\end{equation}
where the charge $Q_\xi$ obtained in this way is the result of the spacetime symmetry described by the vector field $\xi$ in spacetime, $\lrcorner$ represents the contraction of the vector field $\delta_\xi$ with $\Omega$, $\delta_\xi$ is the vector field in phase space generated by the action of the symmetry vector $\xi$, and $\delta$ is the exterior differential. Because of general arguments concerning spatial translations~\cite{JurekAndreaDiscrete}, the canonical non-deformed contraction rule $\delta_\xi \lrcorner \, A\wedge B = (\delta_\xi A) B - A (\delta_\xi B)$ was deformed into
\begin{equation}\label{deformed_contraction}
    \delta_\xi \lrcorner \, A\wedge B = (\delta_\xi A) B + A (S(\delta_\xi) B)
\end{equation}
where $S(\delta \xi) B$ is the same as $\delta_\xi B$ with every momentum switched for its antipode\footnote{Depending on the field $\delta_\xi$, an additional canonical minus sign may be needed besides switching every momentum for its antipode, see~\cite{JurekAndreaDiscrete} for more details.}. 
It is right here where the features pointed out in the previous sections are hidden. In fact, it is possible to show that the property (\ref{deformed_contraction}), introduced for the consistency of the covariant phase space approach, amounts in the end to switching the noncommutative parameter $\epsilon^A$ to the right (only) for the part of the Lagrangian $\cal L(\phi,\phi^\dagger)$ of (\ref{eq-lc}). In turn, this implies that a total derivative (a boundary term) is added to the Lagrangian, precisely the boundary term needed to turn~\eqref{eq-lc} into~\eqref{action_rewriting}. And indeed, the charges obtained from the action \eqref{eq-lc} with this method are exactly the charges in \eqref{charges_cps} (and their counterparts for rotations and boosts), which close the Poincar\'e algebra. In other words, the additional surface term is implicitly considered in the choice \eqref{deformed_contraction}, which allows one to go directly from \eqref{eq-lc} to \eqref{charges_cps}, without needing to rewrite the action as in \eqref{action_rewriting}. 

Summarizing, switching from a Lorentz-invariant to a $\mathcal{C}$-invariant Lagrangian requires introducing different boundary terms. The direct Noether approach, which obtains the charges directly from the boundary term, naturally keeps track of this. On the other hand, such direct computations are already very challenging even for simple models. The covariant phase space approach, while much simpler and transparent from a computational point of view, may hide implicit choices of boundary terms which need to be taken into consideration. In the complex scalar field considered here and in \cite{kDiscrete1,JurekAndreaDiscrete}, it is clear that the relation \eqref{action_rewriting} implies that Lorentz and $\mathcal{C}$ invariance are incompatible\footnote{At least when considering $\mathcal{C}$ transformations which relate $\phi$ and $\phi^\dag$ as in eq.~\eqref{eq-c-def}.}. Generalizing to arbitrary $\kappa$-deformed models, we argue that starting from a Lorentz-invariant Lagrangian, the asymmetry between positive and negative energy modes imposes charges where particles and antiparticles are associated to different domains in momentum space. This in turn translates into different momentum-dependent functions appearing in the $a, a^\dag$ and $b, b^\dag$ sectors in the charges and the symplectic form. Such additional terms allow the charges to close the Poincar\'e algebra (consistently with the Lorentz invariance of the action), but spoil $\mathcal{C}$ invariance. Alternatively, starting from a manifestly $\mathcal{C}$-invariant Lagrangian allows one to obtain manifestly $\mathcal{C}$-invariant charges and symplectic form, which spoil Lorentz (and Poincar\'e) invariance. The relation between the two formulations is a boundary term.
It is important to notice that the latter model, where $\cal C$-invariance is maintained but the charges are not Poincar\'e invariant, can be described only within the canonical approach, since the covariant phase space approach automatically modifies the starting Lagrangian by a boundary term that restores Poincar\'e invariance (but spoils $\cal C$ invariance).

We conjecture that the incompatibility between $\mathcal{C}$ and Lorentz symmetry, shown for the complex $\kappa$-deformed scalar, holds for general $\kappa$-deformed models because of the momentum space geometry.

\section{Parity, time reversal and CPT}
\label{sec:parity and time reversal}

After having identified the origin of the incompatibility between $\cal C$ and Poincar\'e invariance, we describe in this section our new proposal for the remaining discrete symmetry transformations -- parity, time reversal, and finally the combined ${\cal C}{\cal P}{\cal T}$ transformation.
In particular we find that there is a natural choice for the action of time reversal that restores the overall ${\cal C}{\cal P}{\cal T}$ invariance of the Poincar\'e invariant model.

We remark again that there is a certain degree of arbitrariness in the definition of the action of $\kappa$-deformed discrete symmetries. The definitions we give here differ, for instance, from the ones in~\cite{JurekMicheleDiscrete}, which were based on considerations involving the algebra generators -- particularly translations -- on particle states.
Our definition relies solely on the properties of the $\kappa$-field theory defined in sec.~\ref{sec:complex field} and of its momentum space (sec.~\ref{sec:momentum space}).
For the purposes of this work, it suffices to show that there is a possible, well-motivated choice of these definitions that allows us to recover ${\cal C}{\cal P}{\cal T}$-invariance.
We postpone to future studies a thorough critical analysis of the differences in these definitions.

We stress, however, how our new proposal is rather appealing from a relativistic point of view, restoring the link between ${\cal C}{\cal P}{\cal T}$ and Poincar\'e invariance, a pillar of (special) relativity, also in a deformed relativistic setting (we briefly comment on this point in the conclusions).

\subsection{Definition of parity}
\label{parity}

Considering the structure of $\kappa$-momentum space described in sec.~\ref{sec:momentum space}, we expect the theory to be symmetric if one changes the sign of the spatial momentum.
This motivates us to define the parity transformation ${\cal P}$ in the standard way, i.e. as a reflection of the spatial momentum (corresponding to a reflection with respect to the $p_1$  axis in Fig.~\ref{fig:-momentum-space}. See also Fig.~\ref{fig:-momentum-spacePT}):
\begin{equation}
\begin{gathered}
a_{{\bf p}} \stackrel{{\cal P}}{\rightarrow}a_{-{\bf p}}\ , \qquad
b_{\bf p} \stackrel{{\cal P}}{\rightarrow} b_{-{\bf p}}\\
\text{or}\quad \tilde{\phi}\left(\omega_{{\bf p}},{\bf p}\right)\stackrel{{\cal P}}{\rightarrow}\tilde{\phi}\left(\omega_{{\bf p}},-{\bf p}\right)\ .
\end{gathered}
\end{equation}
This was the choice that was also made in the previous works~\cite{kDiscrete1,JurekAndreaDiscrete}.

Considering the Fourier transformed on-shell field expression~\ref{on-shell field} (or~\ref{eq-mode-phi}), it is quite straightforward to show that, substituting the action of {\cal P} and changing variable of integration as ${\bf p}\rightarrow-{\bf p}$
and $S\left({\bf p}\right)\rightarrow-S\left({\bf p}\right)$, one simply obtains
\begin{equation}
\phi\left(t,{\bf x}\right)\stackrel{{\cal P}}{\rightarrow}\phi\left(t,-{\bf x}\right)\ ,
\end{equation}
as in the standard commutative field theory.

\subsection{Definition of time reversal}
\label{time reversal}

Due to the asymmetry of $\kappa$-momentum space, the action of time reversal ${\cal T}$ is more subtle, and requires careful treatment.
We begin by noticing that in the commutative case, besides reversing the direction of (spatial) momenta, ${\bf p}\rightarrow-{\bf p}$,
${\cal T}$ is an antilinear operator (in the quantum setting). In the commutative case, the
antilinearity of the ${\cal T}$ operator is expressed, for the Fourier transformed field, by its action of complex conjugation on the plane wave associated with each mode. There is a subtlety in generalizing this property to the noncommutative case, in that the plane wave is not just a complex number, but an ordered function of noncommutative coordinates, that more precisely is an element of the $AN(3)$ group, which is a group arising in the Iwasawa decomposition ($KAN$ decomposition, see \cite{Helgason}) of $SO(4,1)$ \cite{Arzano:2022ewc}.
One of the consequences is that the complex conjugate of the plane wave involves a transposition of the elements, amounting to an inversion of the group element:
\begin{equation}
\hat{e}_{k}^{\dagger}\left(\hat{x}\right)=e^{-ik_{0}\hat{x}^{0}}e^{-ik_{j}\hat{x}^{j}}=e^{iS\left(k_{j}\right)\hat{x}^{j}}e^{iS\left(k_{0}\right)\hat{x}^{0}}\ ,
\end{equation}
which, under our choice of Weyl map, means that
\begin{equation}
\left(e^{ipx}\right)^{\dagger}=e^{iS\left(p\right)x}\ .
\end{equation}

If we extend this property to the action of ${\cal T}$ on the Fourier transformed field, so that, besides reversing the direction of (spatial) momenta, it gives the complex conjugate of the plane wave associated with each mode, we obtain that the field transforms as
\begin{equation}
\phi\left(x\right)\stackrel{{\cal T}}{\rightarrow}\int\frac{d^{3}p}{2\omega_{{\bf p}}p_{4}/\kappa}\left(a_{-{\bf p}}e^{iS\left(p\right)x}+b_{-{\bf p}}^{\dagger}e^{ipx}\right)\ .\label{TonField_ab}
\end{equation}
By changing again variables of integration as ${\bf p}\rightarrow-{\bf p}$ and $S\left({\bf p}\right)\rightarrow-S\left({\bf p}\right)$, the last expression can be rewritten as\footnote{Notice that in noncommutative coordinates, i.e. before applying the Weyl map, the action of ${\cal T}$ can be expressed as $\hat{\phi}\left(\hat{x}^0,\hat{x}^j\right)\stackrel{{\cal T}}{\rightarrow}\hat{\phi}^{T}\left(-\hat{x}^0,\hat{x}^j\right)$.}
\begin{equation}
\phi\left(t,{\bf x}\right)\stackrel{{\cal T}}{\rightarrow}\phi^{\dagger*}\left(-t,{\bf x}\right)\ .
\end{equation}
This is clearly a generalization of the action of ${\cal T}$ in the undeformed (commutative) theory. Here, in the definition of $\cal T$, is where our new proposal differs from the one in the previous works~\cite{kDiscrete1,JurekAndreaDiscrete}.

With some calculations, it can be shown that (see App.~\ref{app:proveOfEquation}) one can rewrite the r.h.s. of (\ref{TonField_ab}) as
\begin{equation}
\begin{split}
\int\frac{d^{3}p}{2\omega_{p}p_{4}/\kappa} \Big( & \frac{\kappa^{3}}{p_{-}^{3}}a_{S_{-}\left({\bf p}\right)}e^{i\left(-\omega_{{\bf p}}t-{\bf p}\cdot{\bf x}\right)} \\ &
+\frac{\kappa^{3}}{p_{+}^{3}}b_{S_{+}\left({\bf p}\right)}^{\dagger}e^{i\left(\omega_{S\left({\bf p}\right)}t+S_{+}\left({\bf p}\right)\cdot{\bf x}\right)}\Big)\ ,\end{split}
\label{TonField_ab_Sp}
\end{equation}
where $S_{-}({\bf p})$ and $S_{+}({\bf p})$, are the on-shell antipode of the spatial momenta defined in App.~\ref{app:antipodeOnShell}.

\begin{figure}[h]
\hspace{-1cm}
\begin{centering}
\includegraphics[scale=0.28]
{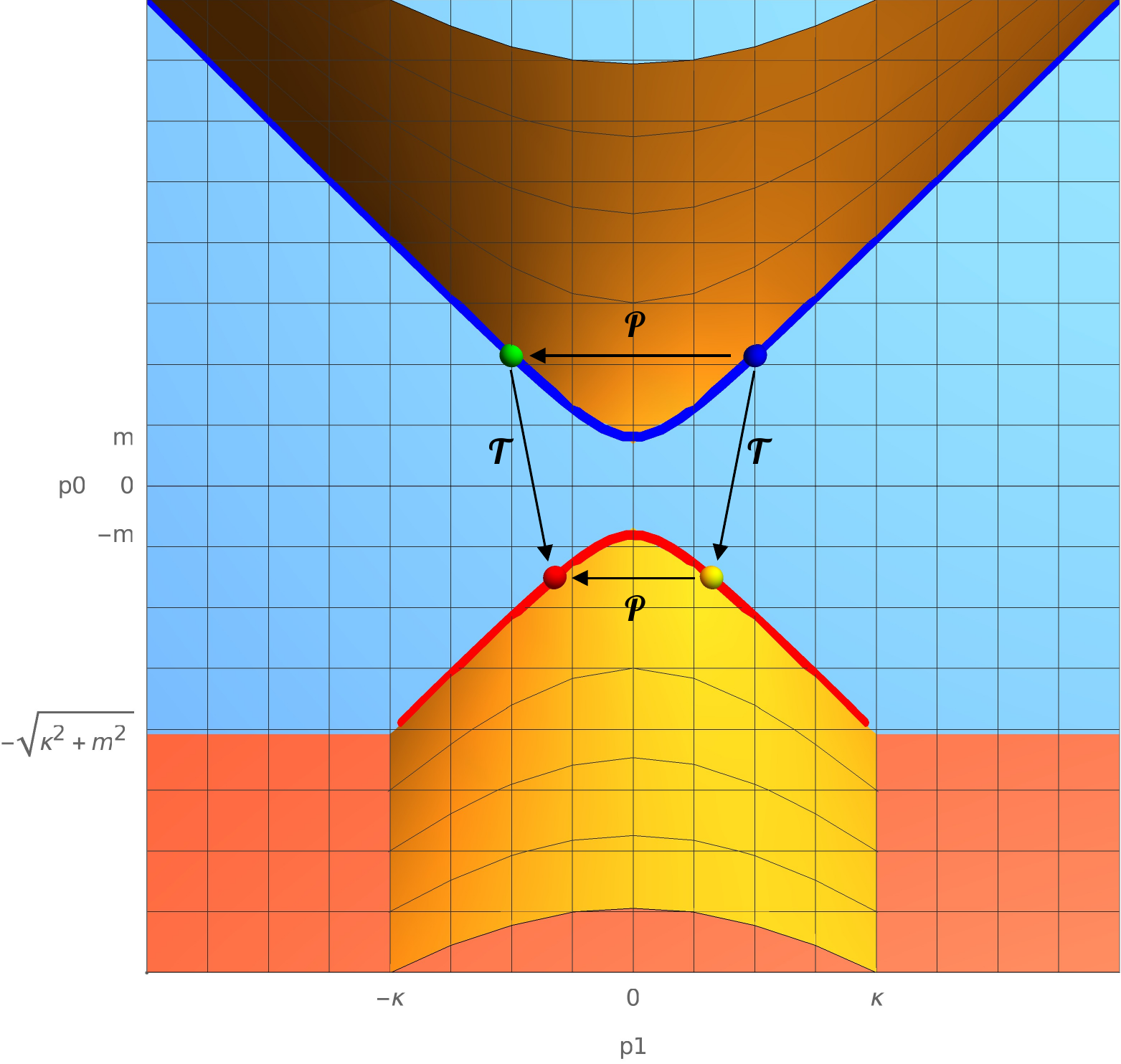}
\caption{\label{fig:-momentum-spacePT}A pictorial description of the action of ${\cal P}$ and ${\cal T}$ in the momentum space of a field.}
\par\end{centering}
\end{figure}

This last expression shows the interesting property that the action
of ${\cal T}$ on fields can be interpreted formally also as simultaneously reversing the sign of the energy of the field modes and changing their spatial momenta into their (on-shell) antipodes.
This property looks intuitively consistent with the picture of $\kappa$-momentum space:
if we look at Fig.~\ref{fig:-momentum-space}, eq.~(\ref{TonField_ab_Sp}) implies that time reversal corresponds to a reflection with respect to the $p_0$ axis, and a replacement of the momenta of the modes with their (on-shell) antipodes (see also Fig.~\ref{fig:-momentum-spacePT}).

\subsection{Action of ${\cal P}{\cal T}$}

If we combine the action of ${\cal P}$ and ${\cal T}$,
we obtain that ${\cal P}{\cal T}$ acts on a noncommutative field
so that
\begin{equation}
\phi\left(x\right)\stackrel{{\cal P}{\cal T}}{\rightarrow}\int\frac{d^{3}p}{2\omega_{{\bf p}}p_{4}/\kappa}\,\left(a_{{\bf p}}e^{iS\left(p\right)x}+b_{{\bf p}}^{\dagger}e^{ipx}\right)\ .
\end{equation}
This rule of action can be written concisely as\footnote{Or, writing it for the noncommutative field, as $\hat{\phi}\left(\hat{x}\right)\stackrel{{\cal P}{\cal T}}{\rightarrow}\hat{\phi}^{T}\left(-\hat{x}\right)$.}
\begin{equation}
\phi\left(x\right)\stackrel{{\cal P}{\cal T}}{\rightarrow}\phi^{\dagger*}\left(-x\right)\ .
\end{equation}

\subsection{CPT transformation}

Finally, let us consider the joint action of ${\cal C}{\cal P}{\cal T}$.
From the previous results, it is straightforward to obtain that the
field changes as
\begin{equation}
\label{CPTphi(x)}
\phi\left(x\right)\stackrel{{\cal C}{\cal P}{\cal T}}{\rightarrow}\phi^{*}\left(-x\right)\ ,
\end{equation}
exactly like for the standard commutative case\footnote{Or, extending the action to the noncommutative field as
$\hat{\phi}\left(\hat{x}\right)\stackrel{{\cal C}{\cal P}{\cal T}}{\rightarrow}\hat{\phi}^{*}\left(-\hat{x}\right)$.}.

Repeating steps similar to the ones for ${\cal T}$, we can give a
more insightful interpretation of the action of ${\cal C}{\cal P}{\cal T}$
on particle/antiparticle modes. Indeed, considering the expression
(\ref{TonField_ab_Sp}), acting again with ${\cal P}$ will change
the sign of ${\bf x}$ (it changes the sign of ${\bf p}$, but we
can reabsorb it into a change of ${\bf x}$), while acting with ${\cal C}$ will switch $a$ modes with $b$ modes. We thus can re-express the change of a field under ${\cal C}{\cal P}{\cal T}$ as
\begin{equation}
\begin{split}
\int\frac{d^{3}p}{2\omega_{p}p_{4}/\kappa}\Big( & \frac{\kappa^{3}}{p_{-}^{3}}b_{S_{-}\left({\bf p}\right)}e^{i\left(-\omega_{{\bf p}}t+{\bf p}\cdot{\bf x}\right)} \\&
+\frac{\kappa^{3}}{p_{+}^{3}}a_{S_{+}\left({\bf p}\right)}^{\dagger}e^{i\left(\omega_{S_{+}\left({\bf p}\right)}t-S_{+}\left({\bf p}\right)\cdot{\bf x}\right)}\Big)\ .
\end{split}
\end{equation}
If we neglect the factors $\kappa^{3}/p_{\pm}^{3}$, the last expression suggests that if a theory is \mbox{${\cal C}{\cal P}{\cal T}$}-invariant, it means it is invariant under a symmetry that exchanges particles with antiparticles traveling backwards in time with antipodal momentum.

\section{CPT invariance}
\label{sec:CPTinvariance}

In this section we finally prove that the action and the charges that we have obtained in Sec.~\ref{sec:complex noether} are invariant under the action of the ${\cal C}{\cal P}{\cal T}$ transformation that we defined in the previous section.

\subsection{Invariance of the action}

We can easily prove the ${\cal C}{\cal P}{\cal T}$-invariance of the action if we write it in terms of noncommutative fields, as in eq.~(\ref{eq-S1-noncomm}).
It is enough to notice that under a ${\cal C}{\cal P}{\cal T}$ transformation the action goes into its complex conjugate, without any transposition (i.e. it is not an inversion) of the plane wave:
\begin{equation}
\hat{{\cal S}}\left(\hat{\phi}^{\dagger},\hat{\phi}\right)\stackrel{{\cal C}{\cal P}{\cal T}}{\rightarrow}
\hat{{\cal S}}\left(\hat{\phi}^{\dagger*},\hat{\phi}^*\right)\ .
\label{CPTactionOnAction}
\end{equation}
Then, if the action is real, it is ${\cal C}{\cal P}{\cal T}$-invariant (as in the commutative case).

It is easy to prove this for the off-shell action.
Indeed, notice first that, using the 4-dimensional (noncommutative) Fourier transform and the properties of plane waves, (\ref{eq-s1}) can be rewritten as
\begin{equation}
\begin{split}\hat{{\cal S}}\left(\hat{\phi}^{\dagger},\hat{\phi}\right) = & \int d\mu\left(k\right)\int d\mu\left(q\right)\
\left(k_\mu q^\mu -m^2\right) \\ &
\times \tilde{\phi}^{*}\left(k\right)\tilde{\phi}\left(q\right)\delta^{4}\left(S\left(k\right)\oplus q\right)\\
= & \int d\mu\left(q\right)\int d^{4}k\ \left(k_\mu q^\mu -m^2\right) \\ &
\times \tilde{\phi}^{*}\left(k\right)\tilde{\phi}\left(q\right)\delta^{4}\left(k-q\right)\\
= & \int d\mu\left(q\right) \left(q_\mu q^\mu -m^2\right)
\tilde{\phi}^{*}\left(q\right)\tilde{\phi}\left(q\right)\ ,
\end{split}
\end{equation}
where we used eqs.~(\ref{delta4noncomm}) to resolve the 4-D delta.
It is straightforward to show that, similarly,
\begin{equation}
\hat{{\cal S}}\left(\hat{\phi}^{\dagger^*},\hat{\phi}^*\right) =
\int d\mu\left(q\right) \left(q_\mu q^\mu -m^2\right)
\tilde{\phi}\left(q\right)\tilde{\phi}^{*}\left(q\right) \ .
\end{equation}
The two results coincide.

\subsection{Invariance of the conserved charges}

To prove that the charges are invariant, it is enough to consider that the net action of ${\cal C}{\cal P}{\cal T}$ on the charges~(\ref{eq-pmu-l1}), (\ref{rotation charge}) and~(\ref{boost charge}) amounts to replacing the momentum space fields (i.e. the $a_{\bf p}$ and $b_{\bf p}$ coefficients) with their complex conjugates.
It is straightforward to show that the translation charges~(\ref{eq-pmu-l1}) are ${\cal C}{\cal P}{\cal T}$-invariant.

For the boost charges, it is convenient to rewrite~(\ref{boost charge}) as
\begin{equation}
{\cal N}_{j}=i\int d^{3}p\frac{p_{+}^{3}}{\kappa^{3}}b_{{\bf p}}^{\dagger}\frac{\partial b_{{\bf p}}}{\partial p^{j}}-i\int d^{3}S\left(p\right)\frac{p_{+}^{3}}{\kappa^{3}}a_{{\bf p}}^{\dagger}\frac{\partial a_{{\bf p}}}{\partial S\left(p^{j}\right)}\ .
\end{equation}
We now integrate by parts half of each term, so to obtain
\begin{equation}
\begin{split}{\cal N}_{j}= & \frac{i}{2}\int d^{3}p\frac{p_{+}^{3}}{\kappa^{3}}\left(b_{{\bf p}}^{\dagger}\frac{\partial b_{{\bf p}}}{\partial p^{j}}-\frac{\partial b_{{\bf p}}^{\dagger}}{\partial p^{j}}b_{{\bf p}}\right)\\
 & -\frac{i}{2}\int d^{3}S\left(p\right)\frac{p_{+}^{3}}{\kappa^{3}}\left(a_{{\bf p}}^{\dagger}\frac{\partial a_{{\bf p}}}{\partial S\left(p^{j}\right)}-\frac{\partial a_{{\bf p}}^{\dagger}}{\partial S\left(p^{j}\right)}a_{{\bf p}}\right)\\
 & -\frac{i}{2\kappa^3}\left( \int d^{3}p\ b_{{\bf p}}^{\dagger}b_{{\bf p}}\frac{\partial p_{+}^{3}}{\partial p^{j}} - \int d^{3}S\left(p\right)a_{{\bf p}}^{\dagger}a_{{\bf p}}\frac{\partial p_{+}^{3}}{\partial S\left(p^{j}\right)}\right)
\end{split}
\end{equation}
Notice now that by the rule~(\ref{CPTphi(x)}), it follows that
\begin{equation}
\begin{split}\frac{\partial\tilde{\phi}\left(p\right)}{\partial p_{\mu}}= & -i\int d^{4}x\ \phi\left(x\right)x^{\mu}e^{-ipx} \\&
\stackrel{{\cal C}{\cal P}{\cal T}}{\rightarrow}i\int d^{4}x\ \phi^{*}\left(-x\right)x^{\mu}e^{-ipx}\\
= & -\int d^{4}x\ \phi^{*}\left(x\right)\frac{\partial}{\partial p_{\mu}}e^{ipx}=-\frac{\partial\tilde{\phi}^{*}\left(p\right)}{\partial p_{\mu}}
\end{split}
\end{equation}
Then, using the definitions~(\ref{eq-mode-phid}), it follows that
\begin{equation}
\frac{\partial a_{{\bf p}}}{\partial p^{j}}\stackrel{{\cal C}{\cal P}{\cal T}}{\rightarrow}-\frac{\partial a^{\dagger}_{\bf p}}{\partial p^{j}}\ ,\qquad\frac{\partial b_{{\bf p}}}{\partial p^{j}}\stackrel{{\cal C}{\cal P}{\cal T}}{\rightarrow}-\frac{\partial b^{\dagger}_{\bf p}}{\partial p^{j}}
\end{equation}
Using these relations and replacing $a_{\bf p}$ and $b_{\bf p}$ with their complex conjugate, we obtain that the boost charges are ${\cal C}{\cal P}{\cal T}$-invariant.

The proof of invariance for rotation charges follows analogously.

\section{Discussion and conclusions}
\label{sec:conclusions}

In this work we studied the properties of deformed symmetries in a framework where spacetime is characterized by noncommutativity of $\kappa$-Minkowski type and its symmetries are described by the $\kappa$-Poincar\'e Hopf algebra.
The momentum space of the theory becomes a group manifold, denoted as $\kappa$-momentum space, where the curvature radius corresponds to the deformation scale.

We used as a test model a scalar field theory defined in this framework, relying on a specific representation based on the so-called classical or embedding basis of $\kappa$-momentum space that was also used in previous works~\cite{FreJurekNowkfield2,kDiscrete1,JurekAndreaDiscrete}.
We performed the Noether analysis of the model and we obtained, for the first time, besides the conserved charges associated with translations, the explicit expression of the Lorentz boost charges through canonical Noether formalism. Apart from the intrinsic interest in these computations, they allowed us to bring to light the apparent incompatibility between $\mathcal{C}$ and Lorentz invariance.

The study of deformed symmetries is particularly relevant for quantum gravity research, where the deformation scale is expected to be close to the Planck scale.
In addition to the deformation of the continuous Poincar\'e symmetries, it is also important to understand the properties of the deformed discrete symmetries: charge conjugation $\mathcal{C}$, parity $\mathcal{P}$, and time reversal $\mathcal{T}$.

As a definition of ${\cal C}$, we considered the same one that was already introduced in the previous works~\cite{kDiscrete1}, \cite{JurekAndreaDiscrete}, and \cite{Bevilacqua:2024jpy}, which corresponds to exchanging the particle and antiparticle modes, and amounts to a conjugation of the field, analogously to the standard undeformed case: $\hat{\phi} \stackrel{{\cal C}}{\rightarrow} \hat{\phi}^\dagger$.
The result of this conjugation is that the momenta of the particle and antiparticle are connected by antipode, as one expects from the structure of $\kappa$-momentum space.
In this sense, one can say that the action of ${\cal C}$ is deformed.

Having obtained the explicit expression for the charges in the canonical formalism, we clarified a puzzling aspect that emerged from the previous studies: symmetrizing the Lagrangian in the attempt to define a ${\cal C}$-invariant field model leads to a breaking of Poincar\'e symmetries; on the other hand, the technique of computation of the charges that was performed in~\cite{kDiscrete1},\cite{JurekAndreaDiscrete}, based on a covariant phase space approach, amounts to adding a boundary term to the action that restores Poincar\'e invariance, but leads to a Lagrangian which is no longer ${\cal C}$-invariant.
This result suggests that in $\kappa$-momentum space, due to the asymmetry between its positive and negative frequency sectors, ${\cal C}$ and Poincar\'e invariance are not compatible.

Regarding the remaining discrete transformations, in the previous works~\cite{kDiscrete1}, \cite{JurekAndreaDiscrete}, \cite{Bevilacqua:2024jpy} the action of $\mathcal{P}$ and $\mathcal{T}$ was kept undeformed. This choice was motivated by the fact that the $\kappa$-deformation defining relation
\begin{align}\label{kappaMink}
[\hat x^0, \hat x^i] = \frac{i}{\kappa} \hat x^i
\end{align}
is covariant under the undeformed action of parity and time reversal.
However, this choice inevitably leads to a violation of $\mathcal{CPT}$ symmetry. From a theoretical perspective, a departure from exact $\mathcal{CPT}$ symmetry would naively seem to contradict the Jost and Greenberg theorem (see~\cite{Greenberg:2003nv} and references therein). Yet, as explained in~\cite{JurekAndreaDiscrete}, the proof of this theorem does not hold in the case of $\kappa$-deformed symmetry. On the phenomenological side, the $\kappa$-deformation of $\mathcal{CPT}$ symmetry implies different decay times for highly energetic particles and antiparticles, which could, in principle, be observed in future accelerators \cite{Bevilacqua:2024jpy}.

In this paper, we proposed a different form of deformed discrete symmetries. Keeping the charge conjugation $\mathcal{C}$ deformed as before, we introduced here a deformed time reversal $\mathcal{T}$, while keeping parity $\mathcal{P}$ undeformed. 
We argue that this choice can be motivated by inspection of the algebraic and geometrical properties of $\kappa$-momentum space.
What we find is that the momentum-space deformations of $\mathcal{C}$ and $\mathcal{T}$ compensate each other, and as a result, the overall (deformed) $\mathcal{CPT}$ symmetry is preserved. It can also be shown that this deformation of discrete symmetries, although different from the one considered previously, still preserves the defining relation \eqref{kappaMink}. The reason for this seemingly paradoxical conclusion lies in the fact that the relation \eqref{kappaMink} captures only leading-order information about $1/\kappa$ non-commutativity, and at this order, both definitions agree. It would be of interest to check whether one could generalize the Greenberg and Jost theorems in the case of the new definition of discrete symmetries.

On the phenomenological side, we expect this new deformation to predict an additional, energy-dependent contribution to $\mathcal{PT}$ symmetry violation. However, restoring a (deformed) $\mathcal{CPT}$ symmetry may lead to the conclusion that in this theory there is no difference between the lifetimes of particles and antiparticles, although this statement should be carefully scrutinized.
We leave a careful investigation of these phenomenological implications to future studies.
We mention, however, that even though they have different conceptual and phenomenological implications, as far as we can say, both modifications of discrete symmetries are self-consistently defined, and we leave it to experiment to decide which one (if any) is realized in nature.

Finally, we remark that these features can perhaps be better understood considering the difference in analogous cases between the effects of a deformation (in the sense of DSR) with respect to those of a violation (in the sense of LIV) of relativistic symmetries~\cite{GACreview,COSTreview}.
It is indeed common for these frameworks that some core properties of (special) relativistic symmetries, like for instance the impossibility of having observer-dependent threshold effects in some particular particle processes, are allowed in a LIV scenario, but are still preserved when symmetries are DSR-deformed but not broken~\cite{GACreview,COSTreview}.
It could be that $\mathcal{CPT}$ symmetry, which seems to be a core feature of a (special) relativistic theory, plays a similar role.

\begin{acknowledgments}
This work falls within the
scopes of the EU COST Action CA23130 “Bridging high and low energies in search of quantum gravity (BridgeQG)”, and was supported by a STSM Grant from the same COST Action CA23130.
For AB, JKG, and GR this work was partially supported by funds provided by the National Science Center, project number 2019/33/B/ST2/00050. 
\end{acknowledgments}

\appendix

\section{Properties of $\kappa$-Poincar\'e/$\kappa$-Minkowski noncommutative spacetime}

We review here some properties of $\kappa$-Poincar\'e/$\kappa$-Minkowski noncommutative spacetime in ``embedding'' momentum space coordinates that are useful to reproduce some of the calculations reported in this work.
Most of these properties were already discussed thoroughly in previous papers, so we avoid dwelling too much on the motivations and details of these structures, and limit ourselves to provide the reader with the formulas most useful for this work.
We refer to the previous papers, in particular~\cite{FreJurekNowkfield2} and~\cite{kDiscrete1} for further details.

\subsection{$\kappa$-Poincar\'e Hopf algebra}
\label{app:kPoincare}

Defining the translation operator $k_\mu$ so that its action $\triangleright$ on the time-ordered noncommutative plane waves~(\ref{noncomm plane wave}) is
\begin{equation}
k_\mu \triangleright \hat{e}_k(\hat{x}) := k_\mu \hat{e}_k(\hat{x}) \ ,
\end{equation}
it follows~\cite{AgoGACdanreaHopf2004} that the $\hat{e}_k(\hat{x})$ form a basis for the  bicrossproduct $\kappa$-Poincar\'e Hopf algebra~\cite{MajidRuegg}
\begin{equation}
\begin{gathered}
\left[k_{\mu},k_{\nu}\right]=0, \quad
\left[R_{j},R_{k}\right]=i\epsilon_{jkl}R_{l}, \\
\left[R_{j}, N_{k}\right]=i\epsilon_{jkl}  N_{l}, \quad
\left[N_{j}, N_{k}\right]=-i\epsilon_{jkl}R_{l}, \\
\left[R_{j},k_{0}\right]=0, ~~~
\left[R_{j},k_{k}\right]=i\epsilon_{jkl}k_{l}, ~~~
\left[N_{j},k_{0}\right]=ik_{j},\\
\left[N_{j},k_{k}\right]=i\delta_{jk}\left(\frac{1-e^{-2\ell k_{0}}}{2\ell}+\frac{\ell}{2}\bf{k}^{2}\right)-i\ell k_{j}k_{k}.
\label{k-poincareBicross}
\end{gathered}
\end{equation}
with coproducts
\begin{equation}
\begin{gathered}
\Delta k_{0}=k_{0}\otimes\mathbf{1}+{\bf 1}\otimes k_{0},\\
\Delta k_{j}=k_{j}\otimes 1 +e^{-\ell k_{0}}\otimes k_{j} \ , \\
\Delta R_{j}=R_{j}\otimes\mathbf{1}+{\bf 1}\otimes R_{j},\\
\Delta N_{j}=N_{j}\otimes 1 +e^{-\ell k_{0}}\otimes N_{j}+\ell\epsilon_{jkl}k_{k}\otimes R_{l}\ .
\end{gathered}
\end{equation}
such that, for instance\footnote{We use here Sweedler notation, such that $\Delta a = a_{(1)}\otimes a_{(2)}= \sum_i a^i_{(1)} \otimes a^i_{(2)}$.},
\begin{equation}
\begin{split}
 k_\mu & \triangleright (\hat{e}_k(\hat{x}) \hat{e}_q(\hat{x}))
= k_\mu \triangleright \hat{e}_{(k\oplus q)}(\hat{x}) \\&
= \cdot (\Delta k_\mu (\hat{e}_k(\hat{x}) \otimes \hat{e}_q(\hat{x}))) \\&
= (k_{\mu(1)} \triangleright \hat{e}_k(\hat{x})) (k_{\mu(2)} \triangleright \hat{e}_q(\hat{x}))) \ .
\end{split}
\end{equation}
The algebra admits a quadratic (mass) Casimir
\begin{equation}
{\cal C}=4\kappa^{2}\sinh^{2}\left(\frac{k_{0}}{2\kappa}\right)-e^{k_{0}/\kappa}{\bf k}^{2} \ .
\end{equation}

If one applies the map~(\ref{embbedding map}) then it is easy to see that the generators corresponding the embedding momenta $p_\mu$ close the standard Poincar\'e algebra
\begin{equation}
\begin{gathered}
\left[R_{j},p_{0}\right]=0, \quad
\left[R_{j},p_{k}\right]=i\epsilon_{jkl}p_{l}, \\
\left[N_{j},p_{0}\right]=ip_{j}, \quad
\left[N_{j},p_{k}\right]=i\delta_{jk}p_0.
\end{gathered}
\end{equation}
The coproducts become
\begin{equation}
\begin{gathered}
\Delta p_0 = p_0\otimes\frac{p_+}{\kappa}+\frac{p_j}{p_+}\otimes p_j+\frac{\kappa}{p_+}\otimes p_0 ,\\
\Delta p_j = p_j\otimes\frac{p_+}{\kappa}+1\otimes p_j \ ,
\end{gathered}
\label{coproducts embedding}
\end{equation}
The mass Casimir becomes the same as in the undeformed case
\begin{equation}
{\cal C} = p_0^2 -{\bf p}^2\ ,
\end{equation}
which justifies the choice of mass-shell constraint adopted in this work.
This is sometime called~\cite{LukKosMasClassicalBasis} the ``classical basis'' of $\kappa$-Poincar\'e.

Another important Hopf-algebraic property is the antipode of the generators, such that, for an element $a$ of the algebra (\ref{k-poincareBicross}),
\begin{equation}
S(a_{(1)})a_{(2)} = a_{(1)}S(a_{(2)}) = 0 .
\end{equation}
The antipodes in the bicrossproduct basis are given by
\begin{equation}
\begin{gathered}
S\left(k_{0}\right)=-k_{0},\quad S\left(k_{j}\right)=-e^{k_{0}/\kappa}k_{j}, \quad
S\left(R_{j}\right)=-R_{j}, \\
S\left(N_{j}\right)=-e^{k_{0}/\kappa}N_{j}+\frac{1}{\kappa}\epsilon_{jkl}e^{k_{0}/\kappa}k_{k}R_{l}.
\end{gathered}
\end{equation}
In the classical (embedding) basis the translation antipodes become
\begin{equation}
S(p_0)= -p_{0}+ \frac{{\bf p}^{2}}{p_{+}}, \quad S(p_j) = -\frac{\kappa}{p_{+}}p_{j} , \quad S(p_4) = p_4.
\end{equation}

\subsection{Differential calculus}\label{app:calculus}
We define the noncommutative derivatives $\hat{\partial}_A$ such that they act on the plane waves through the Weyl map \eqref{embeddingWeylMap} as:
\begin{equation}
\begin{split}
& \hat{\partial}_\mu\hat{e}_k(\hat{x})  = \mathcal{W}\left(\partial_\mu e^{ip_\mu(k)x^\mu}\right) \\&
= \mathcal{W}\left(i p_\mu e^{ip_\mu(k) x^\mu}\right)=ip_\mu\hat{e}_k(\hat{x}) \ ,
\end{split}
\end{equation}
and
\begin{equation}
\hat{\partial}_4\hat{e}_k=i(p_4-\kappa)\hat{e}_k
\end{equation}
Consider now the product of two noncommutative plane waves $\hat{e}_p\hat{e}_q=\mathcal{W}(e^{ip_\mu x^\mu})\mathcal{W}(e^{iq_\mu x^\mu})$. The operator $\hat{\partial}_\mu$ acts on such a product via the coproduct rules inherited from the translation generators:
\begin{equation}
\begin{split}
-i\hat{\partial}_\mu\left(\hat{e}_p\hat{e}_q\right)&=p_\mu\triangleright\hat{e}_p\hat{e_q}=(p_{\mu(1)}\triangleright\hat{e}_p)(p_{\mu(2)}\triangleright\hat{e}_q)\\&=(p\oplus q)_\mu \hat{e}_p\hat{e}_q \ ,
\end{split}
\end{equation}
where
\begin{align}
(p\oplus q)_0&=\frac{q_+}{\kappa}p_0+p_+^{-1}\mathbf{p}\cdot\mathbf{q}+\frac{\kappa}{p_+}q_0\label{eq-oplus-0}\\
(p\oplus q)_j&=\frac{q_+}{\kappa}p_j+q_j \ .
\label{eq-oplus-j}
\end{align}
corresponding to the translation coproducts~(\ref{coproducts embedding}).

The partial derivatives thus inherit the Hopf-algebraic structure from the translation generators. We illustrate this for the spatial derivatives $\partial_j$ as an example. Since $e^{ipx}\star e^{iqx}=\mathcal{W}^{-1}\left(\hat{e}_p\hat{e}_q\right)$, we have
\begin{equation}
e^{ipx}\star e^{iqx}=e^{i(p\oplus q)x}
\label{eq-oplus-star}
\end{equation}
The coproduct follows from
\begin{equation}
\begin{split}
\partial_j(e^{ipx}\star e^{iqx})&=\partial_je^{i(p\oplus q)x}=i(p\oplus q)_je^{i(p\oplus q)x}\\
&=i\left(\frac{q_+}{\kappa}p_j+q_j\right)e^{i(p\oplus q)x}\\
&=\partial_j e^{ipx}\star\frac{\Delta_+}{\kappa}e^{iqx}+e^{ipx}\star\partial_je^{iqx},
\end{split}
\end{equation}
where $\Delta_{+}=-i\partial_{0}-i\partial_{4}+\kappa$.

We obtain the antipode of the derivative by identifying
\begin{equation}
    \begin{split}
    iS(p_j)e^{ipx}&=-i\frac{\kappa}{p_+}p_je^{ipx}=\left(-\kappa\Delta_+^{-1}\partial_j\right)e^{ipx}\\
    &\equiv S(\partial_j)e^{ipx} .
    \end{split}
\end{equation}
In this way, we obtain the following properties:
\begin{equation}
\begin{split}
& \partial_0(f(x)\star g(x)) = \partial_0f\star\Delta _+\kappa^{-1}g \\&
~~~ - i\Delta_+^{-1}\partial_jf\star\partial_jg +\kappa\Delta_+^{-1}f\star\partial_0g \ ,
\end{split}
\end{equation}
\begin{equation}
\partial_j(f\star g)=\partial_jf\star\Delta_+\kappa^{-1}g+f\star\partial_jg \ ,
\end{equation}
\begin{align}
S(\partial_0)&=-\partial_0-i\Delta_+^{-1}\partial_j\partial_j \ ,\label{eq-sd0}\\
S(\partial_j)&=-\kappa\Delta_+^{-1}\partial_j \ ,\label{eq-sdj}\\
S(\partial_4)&=\partial_4\label{eq-sd4}
\end{align}
These relations are used repeatedly for integrating by parts with respect to the $\star$-product in deriving the Noether charges in Sec.~\ref{sec:complex noether}.
Finally notice also that these relations imply that $\hat{\partial_\mu}^\dagger(\hat{\phi})={\cal W}(S(\partial_\mu)\phi)$.

\subsection{Properties of the noncommutative translation and Lorentz parameters}
\label{app:noncomm parameters}

The properties of the noncommutative parameters were derived in~\cite{SitarzDiffCalc} and~\cite{FreJurekNowkfield2} (see also~\cite{FlaviokDiff}).
They can be written in terms of $\star$-products between the differentials and the plane waves as (see also~\cite{kDiscrete1})
\begin{equation}
\begin{gathered}
dx^{A}\star \phi(x) = \left(K\right)_{B}^{A}\left(\partial \right) \phi(x) \star dx^{B}, \\
\omega^{\mu\nu}\star \phi(x) =\left(\Omega^{-1}\right)_{\rho\sigma}^{\mu\nu}\left(\partial\right) \phi(x) \star\omega^{\rho\sigma}
\label{transParStar}
\end{gathered}
\end{equation}
where
\begin{equation}
K\left(p\right)=\frac{1}{\kappa}\left(\begin{array}{ccc}
p_{4}+\frac{{\bf p}^{2}}{p_{0}+p_{4}} & {\bf p}^{T} & -p_{0}+\frac{{\bf p}^{2}}{p_{0}+p_{4}}\\
\frac{\kappa}{p_{0}+p_{4}}{\bf p} & \kappa{\bf 1}_{3\times3} & \frac{\kappa}{p_{0}+p_{4}}{\bf p}\\
-p_{0} & -{\bf p}^{T} & p_{4}
\end{array}\right),
\end{equation}
and
\begin{equation}
\Omega_{\rho\sigma}^{\mu\nu}\left(p\right)=\delta_{[\rho}^{\,\mu}\tau_{\sigma]}^{\,\nu}\left(p\right)
\end{equation}
with
\begin{equation}
\tau\left(p\right)=\left(\begin{array}{cc}
2\frac{\kappa}{p_{0}+p_{4}}-1\quad & -2\frac{{\bf p}}{p_{0}+p_{4}}\\
0 & {\bf 1}
\end{array}\right).
\end{equation}

\subsection{Deformed composition of plane waves}\label{app:plane waves}
Using \eqref{eq-oplus-0} and \eqref{eq-oplus-j}, we can evaluate the integrals over $dx$ of $\star$-composed plane waves as Dirac deltas, e.g.
\begin{equation}
\begin{split}
&\int d^4x d^4p d^4q F(p_0,\mathbf{p},q_0,\mathbf{q})e^{i(p\oplus q)x}\\
=&\;\int d^4x d^4p d^4q F(p_0,\mathbf{p},q_0,\mathbf{q})e^{i(p\oplus q)_0 x^0}e^{i\frac{q_+}{\kappa}\left(p_j-S(q_j)\right)x^j}\\
=&\;\int dx^0 dp_0 d^4q F(p_0,S(\mathbf{q}),q_0,\mathbf{q})\frac{\kappa^3}{q_+^3}e^{i\left(\frac{q_{+}}{\kappa}p_{0}+\frac{\kappa}{p_{+}}S(q_0)\right)x^0}\\
=&\;\int d^4q F(S(q_0),S(\mathbf{q}),q_0,\mathbf{q})\frac{\kappa^3}{q_+^3}\frac{q_4}{\kappa}
\end{split}
\end{equation}
This way we obtain the following identities
\begin{align}
&\int d^4x e^{i(p\oplus q)x}=\frac{\kappa^3}{q_+^3}\frac{q_4}{\kappa}\delta^4(p-S(q))\\
&\int d^4x e^{i(p\oplus S(q))x}=\frac{q_+^3}{\kappa^3}\frac{q_4}{\kappa}\delta^4(p-q)\label{eq-d4-ps}\\
&\int d^4x e^{i(S(p)\oplus q)x}=\frac{\kappa^3}{q_+^3}\frac{q_4}{\kappa}\delta^4(S(p)-S(q))\label{eq-d4-sp}\\
&\int d^4x e^{i(S(p)\oplus S(q))x}=\frac{q_+^3}{\kappa^3}\frac{q_4}{\kappa}\delta^4(S(p)-q)
\end{align}
and their 3-dimensional (on-shell) counterparts:
\begin{align}
&\int d^3x e^{i(p\oplus q)_j x^j}=\frac{\kappa^3}{q_+^3}\delta^3(\mathbf{p}-S(\mathbf{q}))\\
&\int d^3x e^{i(p\oplus S(q))_j x^j}=\frac{q_+^3}{\kappa^3}\delta^3(\mathbf{p}-\mathbf{q})\label{eq-delta3-ps}\\
&\int d^3x e^{i(S(p)\oplus q)_j x^j}=\frac{\kappa^3}{q_+^3}\delta^3(S(\mathbf{p})-S(\mathbf{q}))\\
&\int d^3x e^{i(S(p)\oplus S(q))_j x^j}=\frac{q_+^3}{\kappa^3}\delta^3(S(\mathbf{p})-\mathbf{q})
\end{align}

Similar relations can be obtained easily for the 4-dimensional deltas in terms of the momentum space coordinates $k_\mu$ defining the noncommutative plane wave~(\ref{noncomm plane wave}):
\begin{equation}
\begin{gathered}\delta^{4}\left(k\oplus q\right)=e^{3k_{0}/\kappa}\delta^{4}\left(S\left(k\right)-q\right) \ , \\
\delta^{4}\left(k\oplus S\left(q\right)\right)=e^{3k_{0}/\kappa}\delta^{4}\left(S\left(k\right)-S\left(q\right)\right) \ ,\\
\delta^{4}\left(S\left(k\right)\oplus q\right)=e^{-3k_{0}/\kappa}\delta^{4}\left(k-q\right) \ ,\\
\delta^{4}\left(S\left(k\right)\oplus S\left(q\right)\right)=e^{-3k_{0}/\kappa}\delta^{4}\left(k-S\left(q\right)\right) \ .
\end{gathered}
\label{delta4noncomm}
\end{equation}

\subsection{$\star$-product for $x_\mu$}
\label{app:star x}
The product $x_\mu\star e^{iqx}$ can be expressed as
\begin{align}
\begin{split}
x_\mu\star e^{iqx}&=\left(\left.-i\frac{\partial}{\partial p^\mu}\right|_{p=0}e^{ipx}\right)\star e^{iqx}\\
&=\left.-i\frac{\partial}{\partial p^\mu}\right|_{p=0}e^{i(p\oplus q)x}
\end{split}
\end{align}
Calculating the respective $p_0$ and $p_j$ derivatives of \eqref{eq-oplus-0} and \eqref{eq-oplus-j} and putting $p$ to 0, we obtain
\begin{equation}
x_\mu\star e^{iqx}=\frac{1}{\kappa}\left(x_\mu q_+-x_0q_\mu\right)e^{iqx}
\end{equation}
Now for any function $\phi(x)$ we can write
\begin{equation}
\begin{split}
x_\mu\star\phi(x)&=\int d^4p \tilde\phi(p)x_\mu\star e^{ipx}\\
&=\int d^4p \tilde\phi(p)\frac{1}{\kappa}\left(x_\mu p_+-x_0p_\mu\right)e^{ipx}\\
&=\frac{1}{\kappa}\left(x_\mu\Delta_++ix_0\partial_\mu\right)\phi(x)
\end{split}
\end{equation}

\subsection{Twisted cyclicity}\label{app:twisted}
While the $\star$-product is inherently noncommutative, it possesses \emph{twisted cyclicity} under integration:
\begin{equation}
\int d^4xf\star g=\int d^4 xg\star\left(\frac{\Delta_+^3}{\kappa^3}f\right)=\int d^4 x\left(\frac{\kappa^3}{\Delta_+^3}g\right)\star f
\end{equation}
This can be proved directly from the Fourier transform:
\begin{equation}
\begin{split}
&\int d^{4}xf(x)\star g(x)	\\
=&\;\int d^{4}x\int \frac{d^{4}p}{p_4/\kappa}\int \frac{d^{4}S(q)}{q_4/\kappa}f(p)g(S(q))e^{i(p\oplus S(q))x}\\
	=&\;\int \frac{d^{4}p}{p_4/\kappa}\int d^{4}S(q)f(p)g(S(q))\frac{q_{+}^{3}}{\kappa^{3}}\delta^{4}(p-q)\\
	=&\;\int \frac{d^{4}S(q)}{q_4/\kappa}f(q)g(S(q))\frac{q_{+}^{3}}{\kappa^{3}}=\int \frac{d^{4}q}{q_4/\kappa} f(q)g(S(q))
\end{split}
\end{equation}
while
\begin{equation}
\begin{split}
&\int d^{4}x g(x)\star f(x)	\\
=&\;\int d^{4}x\int \frac{d^{4}S(p)}{p_4/\kappa}\int \frac{d^{4}q}{q_4/\kappa} f(q)g(S(p))e^{i(S(p)\oplus q)x}\\
	=&\;\int \frac{d^{4}S(p)}{p_4/\kappa}\int d^{4}q f(q)g(S(p))\frac{\kappa^{3}}{q_{+}^{3}}\delta^{4}(S(p)-S(q))\\
	=&\;\int \frac{d^{4}q}{q_4/\kappa} f(q)g(S(q))\frac{\kappa^{3}}{q_{+}^{3}}=\int d^{4}x\frac{\kappa^{3}}{\Delta_{+}^{3}}f(x)\star g(x)\\
    =&\;\int d^{4}x f(x)\star\frac{\Delta_{+}^{3}}{\kappa^{3}}g(x)
\end{split}
\end{equation}
where we made use of \eqref{eq-d4-ps} and \eqref{eq-d4-sp}.

\section{Properties of the antipode on-shell}
\label{app:antipodeOnShell}

Since the main results of our work rely on the structure of $\kappa$-momentum space, we think that the properties of the momentum-space antipode on-shell require a more careful description.
These properties were already discussed in~\cite{FreJurekNowkfield2}, and we refer to that paper for further insights.

\subsection{Positive and negative frequency antipodes}

On-shell, we can consider only the antipode of the spatial
momentum. On the positive frequency mass-shell orbit, one has
\begin{equation}
p_{0}=\omega_{{\bf p}}=\sqrt{{\bf p}^{2}+m^{2}}\ ,
\end{equation}
which, under antipode, gives
\begin{equation}
\omega_{S\left({\bf p}\right)}=\omega_{{\bf p}}-p_{+}^{-1}{\bf p}^{2}=-S\left(p_{0}\right)\Big|_{p_{0}=\omega_{{\bf p}}}\ .\label{omegaSp}
\end{equation}
While the four-dimensional antipode is an involution that maps the positive frequency mass-shell to
the negative and vice-versa, from the last relation
we see that if we act with only the spatial antipode we are still
in the positive frequency mass-shell. It follows that the spatial
antipode, by itself, is not an involution, since
\begin{equation}
S\left(S\left({\bf p}\right)\right)={\bf p}+\frac{2{\bf p}^{2}}{\kappa^{2}-2{\bf p}^{2}}{\bf p}\neq{\bf p}\ .
\end{equation}

Following~\cite{FreJurekNowkfield2}, one can however recover a ``semi-involution
property'' considering the role of the negative frequencies. Let
us define $p_{-}\left({\bf p}\right)$ as (here and in the following
all variables are to be considered as functions of ${\bf p}$)
\begin{equation}
p_{-}=p_{+}\Big|_{p_{0}=-\omega_{{\bf p}}}=p_{4}+p_{0}\Big|_{p_{0}=-\omega_{{\bf p}}}=p_{4}-\omega_{{\bf p}}\ .\label{pminus}
\end{equation}
Let us now distinguish between the spatial antipode on the positive
shell and the one on the negative shell. Let us call them respecively\footnote{Notice the change of sign, so that $S_{+}\left({\bf p}\right)$
and $S_{-}\left({\bf p}\right)$ have the same sign of ${\bf p}$.}
\begin{equation}
S_{+}\left({\bf p}\right)=-S\left({\bf p}\right)\Big|_{p_{0}=\omega_{{\bf p}}}=\frac{\kappa}{p_{+}}{\bf p}=\frac{\kappa}{p_{4}+\omega_{{\bf p}}}{\bf p},\label{S+}
\end{equation}
and
\begin{equation}
S_{-}\left({\bf p}\right)=-S\left({\bf p}\right)\Big|_{p_{0}=-\omega_{{\bf p}}}=\frac{\kappa}{p_{-}}{\bf p}=\frac{\kappa}{p_{4}-\omega_{{\bf p}}}{\bf p}\ .\label{S-}
\end{equation}
From (\ref{omegaSp}) we have that
\begin{equation}
\omega_{S_{+}\left({\bf p}\right)}=\omega_{{\bf p}}-p_{+}^{-1}{\bf p}^{2}\ .
\end{equation}
 It is easy to show that
\begin{equation}
\omega_{S_{-}\left({\bf p}\right)}=\omega_{{\bf p}}+p_{-}^{-1}{\bf p}^{2}=S\left(p_{0}\right)\Big|_{p_{0}=-\omega_{{\bf p}}}\label{omegaS-p}
\end{equation}
We find also the relations
\begin{equation}
p_{+}\left(S^{-}\left({\bf p}\right)\right)=p_{4}+\omega_{S^{-}\left({\bf p}\right)}=\frac{\kappa^{2}}{p_{-}}\ ,\label{S-p+}
\end{equation}
and
\begin{equation}
p_{-}\left(S_{+}\left({\bf p}\right)\right)=p_{4}-\omega_{S^{+}\left({\bf p}\right)}=\frac{\kappa^{2}}{p_{+}}\ ,\label{S+p-}
\end{equation}
From these relations it follows also that
\begin{equation}
S^{-}\left(S^{+}\left({\bf p}\right)\right)=S^{+}\left(S^{-}\left({\bf p}\right)\right)={\bf p}\ ,\label{semiInvolution}
\end{equation}

\subsection{Change of the spatial integration measure under antipode}

\label{app:AntipodeMeasure}

Under the change ${\bf p}\rightarrow S\left({\bf p}\right)$, the
integration measure changes as\footnote{Since the domain of integration of $S\left({\bf p}\right)$ is bounded
to the values $\left|S\left({\bf p}\right)\right|<\kappa$, the overall
change of integral is
\begin{equation}
\int\frac{d^{3}p}{\omega_{{\bf p}}}\ \frac{\kappa^{3}}{p_{+}^{3}}=\int_{\left|S_{+}\left({\bf p}\right)\right|<\kappa}\frac{d^{3}S_{+}\left(p\right)}{\omega_{S_{+}\left({\bf p}\right)}}.
\end{equation}
}
\begin{equation}
d^{3}S\left({\bf p}\right)=d^{3}p\ \frac{\kappa^{3}}{p^{3}}\frac{\omega_{S\left({\bf p}\right)}}{\omega_{{\bf p}}}\ .
\end{equation}

\newpage

\onecolumngrid

\noindent
More generally, one can prove that
\begin{equation}
d^{3}S_{\pm}\left(p\right)=d^{3}p\ \frac{\kappa^{3}}{p_{\pm}^{3}}\frac{\omega_{S_{\pm}\left({\bf p}\right)}}{\omega_{{\bf p}}}\ .\label{measureAntipode}
\end{equation}

\section{Proof of the expression (\ref{TonField_ab_Sp})}
\label{app:proveOfEquation}

Consider the action of ${\cal T}$ on fields (\ref{TonField_ab}),
that we can rewrite explicitly as
\begin{equation*}
\phi\left(x\right)\stackrel{{\cal T}}{\rightarrow}\int\frac{d^{3}p}{2\omega_{{\bf p}}p_{4}/\kappa}\left(a_{-{\bf p}}e^{i\left(-\omega_{S\left({\bf p}\right)}t-S\left({\bf p}\right)\cdot{\bf x}\right)}+b_{-{\bf p}}^{\dagger}e^{i\left(\omega_{{\bf p}}t-{\bf p}\cdot{\bf x}\right)}\right)\ .
\end{equation*}
Using the properties obtained in the last sections, we can rewrite
it as
\[
\begin{split} & \int\frac{d^{3}S_{-}\left(S_{+}\left(p\right)\right)}{2\omega_{S_{-}\left(S_{+}\left(p\right)\right)}p_{4}/\kappa}a_{-S_{-}\left(S_{+}\left({\bf p}\right)\right)}e^{i\left(-\omega_{S_{+}\left({\bf p}\right)}t+S_{+}\left({\bf p}\right)\cdot{\bf x}\right)} \\&
\!\!\!\! + \!\! \int\frac{d^{3}S_{+}\left(S_{-}\left({\bf p}\right)\right)}{2\omega_{S_{+}\left(S_{-}\left({\bf p}\right)\right)}p_{4}/\kappa}b_{-S_{+}\left(S_{-}\left({\bf p}\right)\right)}^{\dagger}e^{i\left(\omega_{S_{+}\left(S_{-}\left({\bf p}\right)\right)}t-S_{+}\left(S_{-}\left({\bf p}\right)\right)\cdot{\bf x}\right)}\\
= & \int\frac{d^{3}S_{+}\left(p\right)}{2\omega_{S_{+}\left(p\right)}p_{4}/\kappa}\frac{\kappa^{3}}{p_{-}^{3}\left(S_{+}\left({\bf p}\right)\right)}a_{-S_{-}\left(S_{+}\left({\bf p}\right)\right)}e^{i\left(-\omega_{S_{+}\left({\bf p}\right)}t+S_{+}\left({\bf p}\right)\cdot{\bf x}\right)}\\ &
\hspace{-0.5cm} + \!\! \int \!\!\! \frac{d^{3}S_{-}\left({\bf p}\right)}{2\omega_{S_{-}\left({\bf p}\right)}p_{4}/\kappa}\frac{\kappa^{3}}{p_{+}^{3}\left(S_{-}\left({\bf p}\right)\right)}b_{-S_{+}\left(S_{-}\left({\bf p}\right)\right)}^{\dagger}e^{i\left(\omega_{S_{+}\left(S_{-}\left({\bf p}\right)\right)}t-S_{+}\left(S_{-}\left({\bf p}\right)\right)\cdot{\bf x}\right)}\\
& \hspace{-0.7cm} = \!\!  \int \!\!\! \frac{d^{3}p}{2\omega_{{\bf p}}p_{4}/\kappa}\! \left(\frac{\kappa^{3}}{p_{-}^{3}}a_{S_{-}\left({\bf p}\right)}e^{-i\left(\omega_{{\bf p}}t+{\bf p}\cdot{\bf x}\right)} \!+\! \frac{\kappa^{3}}{p_{+}^{3}}b_{S_{+}\left({\bf p}\right)}^{\dagger}e^{i\left(\omega_{S_{+}\left({\bf p}\right)}t+S_{+}\left({\bf p}\right)\cdot{\bf x}\right)}\!\right)
\end{split}
\]
where in the last passage we substituted $S_{+}\left({\bf p}\right)\rightarrow-{\bf p}$
in the first integral and $S_{-}\left({\bf p}\right)\rightarrow-{\bf p}$
in the second integral.

%\newpage
%\phantom{a}
%\newpage

%\vspace{0.5 cm}

\twocolumngrid

\end{document}